\documentclass[prb]{revtex4}
\usepackage{amsmath,amssymb,latexsym,epsfig,graphics,epsf,bm}
\usepackage{graphics}
\usepackage{float}
\usepackage{placeins}
\newcommand{\br}{\mathbf r}

\newcommand{\bR}{\mathbf R}

\newcommand{\be}{\begin{equation}}
\newcommand{\ee}{\end{equation}}

\newcommand{\sex}{{s}_{\rm ex}}

\newcommand{\tU}{\tilde{\Phi}}
\newcommand{\bRa}{{\bf R}_{\rm a}}
\newcommand{\bRb}{{\bf R}_{\rm b}}
\renewcommand{\o}{\char 28}
\usepackage{color}

\begin{document}

\title{Isomorph invariance of classical crystals' structure and dynamics}
\author{Dan E. Albrechtsen, Andreas E. Olsen, Ulf R. Pedersen, Thomas B. Schr{\o}der, and Jeppe C. Dyre}
\affiliation{DNRF Center ``Glass and Time'', IMFUFA, Dept. of Sciences, Roskilde University, P. O. Box 260, DK-4000 Roskilde, Denmark}
\date{\today}

\begin{abstract}
This paper shows by computer simulations that some crystalline systems have curves in their thermodynamic phase diagrams, so-called isomorphs, along which structure and dynamics in reduced units are invariant to a good approximation. The crystals are studied in a classical-mechanical framework, which is generally a good description except significantly below melting. The existence of isomorphs for crystals is validated by simulations of particles interacting via the Lennard-Jones pair potential arranged into a face-centered cubic (FCC) crystalline structure; the slow vacancy-jump dynamics of a defective FCC crystal is also shown to be isomorph invariant. In contrast, a NaCl crystal model does not exhibit isomorph invariances. Other systems simulated, though in less detail, are the Wahnstr{\"o}m binary Lennard-Jones crystal with the ${\rm MgZn_2}$ Laves crystal structure, monatomic FCC crystals of particles interacting via the Buckingham pair potential and via a novel purely repulsive pair potential diverging at a finite separation, an ortho-terphenyl molecular model, and SPC/E hexagonal ice. Except for NaCl and ice, the crystals simulated all have isomorphs. Based on these findings and previous simulations of liquid models, we conjecture that crystalline solids with isomorphs include most or all formed by atoms or molecules interacting via metallic or van der Waals forces, whereas covalently- or hydrogen-bonded crystals are not expected to have isomorphs. Crystals of ions or dipolar molecules constitute a limiting case for which isomorphs are only expected when the Coulomb interactions are relatively weak. We briefly discuss the consequences of the findings for theories of melting and crystallization.
\end{abstract}
\maketitle

\section{Introduction}

In many situations the physical properties of crystals are properly described by classical mechanics \cite{ash76,kit76,sidebottom}. Of course, the fact that the specific heat vanishes when temperature is lowered towards zero can only be explained by invoking quantization of the phonon field, but otherwise it makes good sense to evaluate a crystal's structural and dynamic properties from a purely classical description. This is obviously the case for crystals of large particles like those of colloids \cite{bia12,jua12} or dust plasmas \cite{lam00,dur13}, but crystals of atoms and molecules not significantly below the melting temperature are also generally well described by classical mechanics \cite{ubb65,boc13}. Thus current theories and computer simulations of melting, superheated crystals, etc, are all formulated within a classical, Newtonian framework \cite{for05,bai08a,wan08,gal10,zha11,ped13}. 

This paper presents computer simulations of crystals of classical-mechanical particles with a focus on the standard Lennard-Jones (LJ) system, demonstrating invariance of structure and dynamics to a very good approximation along the configurational adiabats. Recall that if $r$ is the distance between two particles, the LJ pair potential is defined \cite{lj24} by

\be 
v_{\rm LJ}(r)=4\varepsilon\left[\left(\frac{\sigma}{r}\right)^{12}-\left(\frac{\sigma}{r}\right)^{6}\right]\,.
\ee
This potential diverges as $r\rightarrow 0$ and goes to zero as $r\rightarrow\infty$; the prefactor $4\varepsilon$ ensures that the minimum value of $v(r)$ is $-\varepsilon$. For many years the LJ potential has been the standard pair potential for theoretical studies of liquids \cite{and05,eav09,han13}. It is also the most widely computer-simulated potential, because it has become a standard building block in molecular models and models of complex systems like biomembranes and large biomolecules \cite{mac04,stone}.

The theory of isomorphs, which was proposed for liquids in Ref.~\cite{IV}, predicts that the configurational adiabats in the thermodynamic phase diagram of certain systems are curves of invariance for many properties. The main finding of the present paper is that many solids also have isomorphs; in fact it turns out that the isomorph theory works even better here than for liquids. The theory does not work well for crystalline sodium chloride or for ice, however, which shows that the existence of isomorphs is not a trivial consequence of crystallinity.

Although more simulations are needed, the picture that emerges from the results reported below  is that any liquid with isomorphs also has isomorphs in the crystalline phase. Thus ``having isomorphs'' is a material-specific property that survives a first-order phase transition, an unusual situation in condensed matter physics.

The invariances found along isomorphs have several important consequences. For instance, the fact that the melting curve is itself an isomorph \cite{IV,V} implies invariance along this curve of reduced-unit vibrational mean-square displacement, configurational entropy, isochoric specific heat, radial distribution function(s), phonon spectrum, etc. Such melting-line invariances have been known for many years from simulations and experiments \cite{ubb65,bor39,tal80,wallace}, but only now get a unified theoretical explanation. In fact, melting line invariants referring to a single phase -- liquid or solid -- have always presented a challenge to theory because the melting line is where the two phases' free energies are identical: how can one phase know about the free energy of another? The isomorph theory as validated below for crystalline systems offers a resolution of this paradox for the systems with isomorphs, a class now known as ``Roskilde-simple'' or just ``Roskilde'' systems \cite{fle14,pra14,pie14}. Further isomorph-theory predictions is that the dynamics of melting, crystal nucleation, and crystal growth in properly reduced units are all invariant along the isomorphs. For a fixed degree of superheating/supercooling this implies temperature/pressure independence of the melting and crystallization rates and mechanisms. 

The paper is structured as follows. In Sec. \ref{sim} we give a few details of the computer simulations carried out. Section \ref{poten} presents data for the potential energies of scaled versus original configurations, showing a clear difference in the behavior between the LJ crystal and a NaCl model crystal. For the LJ crystal the results allow one to identify isomorphic state points, and Sec. \ref{structure} shows data validating the predicted isomorph invariance of structure and dynamics along an isomorph; in contrast, the NaCl crystal does not exhibit these invariances. Section \ref{jump} shows that jumps along an isomorph for the LJ crystal lead to instantaneous thermal equilibration. The findings for the LJ and the NaCl crystals are put into a broader perspective in Sec. \ref{other}, which presents data for five other systems, three atomic and two molecular crystals. Section \ref{disc} gives a discussion.

\section{Some simulation details}\label{sim}

All simulations were standard Nose-Hoover $NVT$ simulations carried out using the Roskilde University Molecular Dynamics (RUMD) software package \cite{rumd}. Periodic boundary conditions were employed throughout. The first system studied below consists of 4000 LJ particles arranged into a face-centered cubic (FCC) $10\times 10\times 10$ lattice \cite{rumd}. This is the stable crystal structure of the LJ \cite{sti01} and many other simple systems like, e.g., that defined by the Buckingham potential (Sec. \ref{other}). The potential was cut using the shifted-forces method \cite{tildesley} with cut-off distance $3.5\rho^{-1/3}$ independent of the LJ parameter $\sigma$, corresponding to cut-offs larger than $2.6\sigma$ at all densities simulated. 

For a systems of $N$ particles in volume $V$, the density is defined by $\rho\equiv N/V$. The simulations were carried out in the so-called reduced units defined by the length unit $\rho^{-1/3}$, energy unit $k_BT$, and time unit $\rho^{-1/3}\sqrt{m/k_BT}$ in which $m$ is the particle mass. This is useful in order to avoid overflow problems when large density and temperature changes are involved. In practice, the density and temperature were both kept constant equal to unity -- instead the pair potential's energy and length parameters $\varepsilon$ and $\sigma$ were adjusted from state point to state point. The same physics is obtained in this way as by varying density and temperature for fixed $\sigma$ and $\varepsilon$ because structure and dynamics in reduced units depend only on the dimensionless parameters $\rho\sigma^3$ and $k_BT/\varepsilon$. The time step was 0.0025 in reduced units and the time constant of the Nose-Hoover thermostat was kept constant in reduced units. 

The NaCl model studied consisted of $2\cdot 1372$ particles placed in two interpenetrating $7\times 7\times 7$ FCC lattices. The pair potential is a LJ potential plus a Coulomb term \cite{smi94}, 

\be
v_{ij}(r)=4\varepsilon_{ij}\left[\left(\frac{\sigma_{ij}}{r}\right)^{12}-\left(\frac{\sigma_{ij}}{r}\right)^{6}\right]+\frac{z_i z_j e^2}{4\pi\epsilon_0 r}\,.
\ee
Here $\varepsilon_{ij}$, $\sigma_{ij}$, and $z_{i}$ have different values depending on the three possible atom-atom interactions, see Ref. \onlinecite{smi94} for details. Due to the long-ranged nature of the Coulomb part of the NaCl potential \cite{kal11}, a shifted-force cutoff with $r_{\rm cut}=6.5\rho^{-1/3}$ was used \cite{han12}. 

For the simulations of Sec. \ref{other} the crystals were of the following sizes. The Wahnstr{\"o}m binary LJ crystal consisted of 864 small and 432 large particles, the Buckingham and the ``sum-IPL'' crystals were each of 4000 particles, the OTP crystal consisted of 324 molecules, and the ice crystal of 432 molecules.

\section{Potential energies of scaled configurations and the direct isomorph check}\label{poten}

This section motivates the isomorph concept by presenting simulation data for how the potential energy of a uniformly scaled crystalline configuration relates to that of the original configuration (taken from an equilibrium simulation). A recent brief review of the isomorph theory and the evidence for it coming from simulations and experiments is given in Ref. \onlinecite{dyr14}; the theory of Roskilde systems is given in a series of five comprehensive papers \cite{I,II,III,IV,V}.

\begin{figure}
\centering
\includegraphics[width=.3\textwidth]{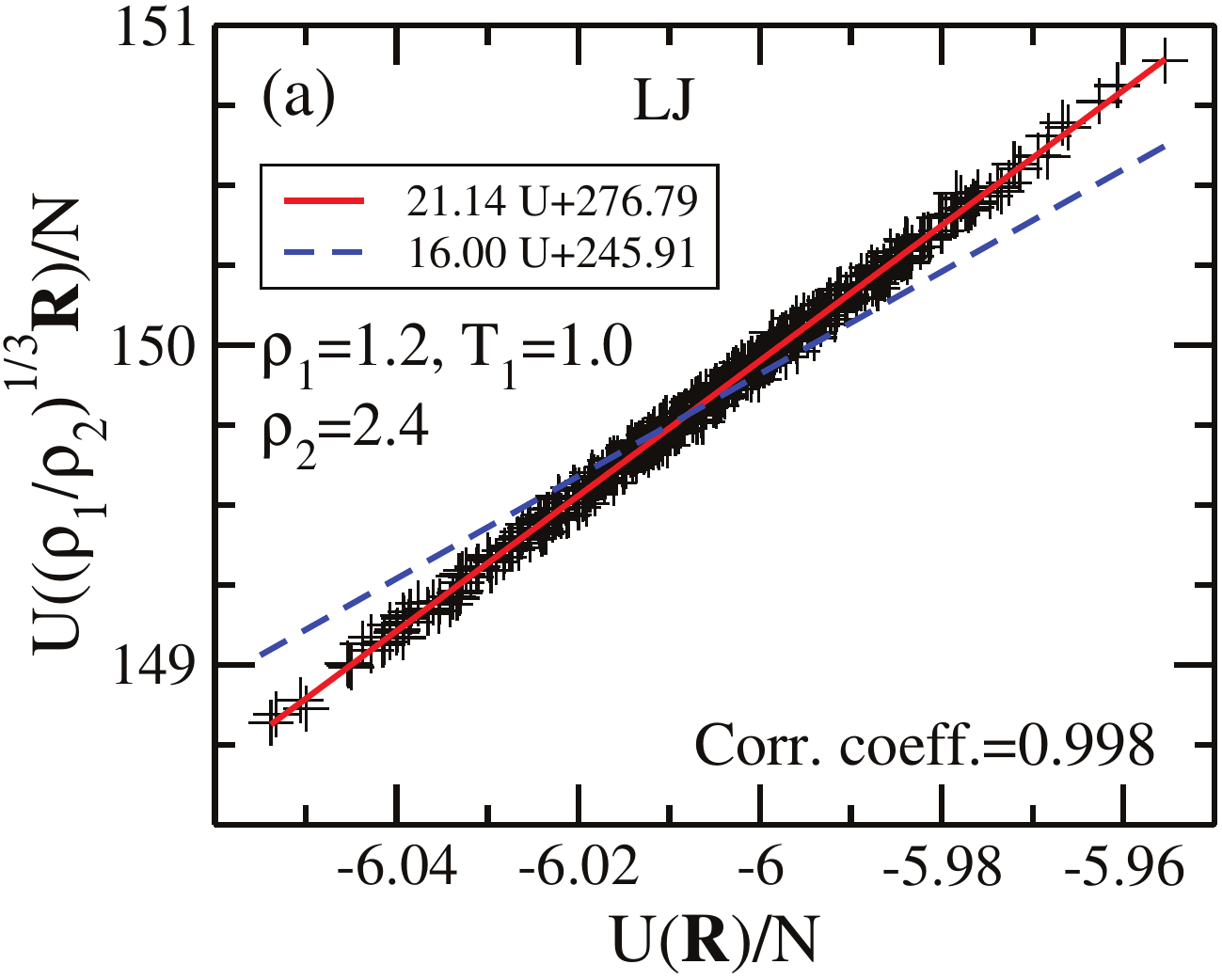}
\includegraphics[width=.3\textwidth]{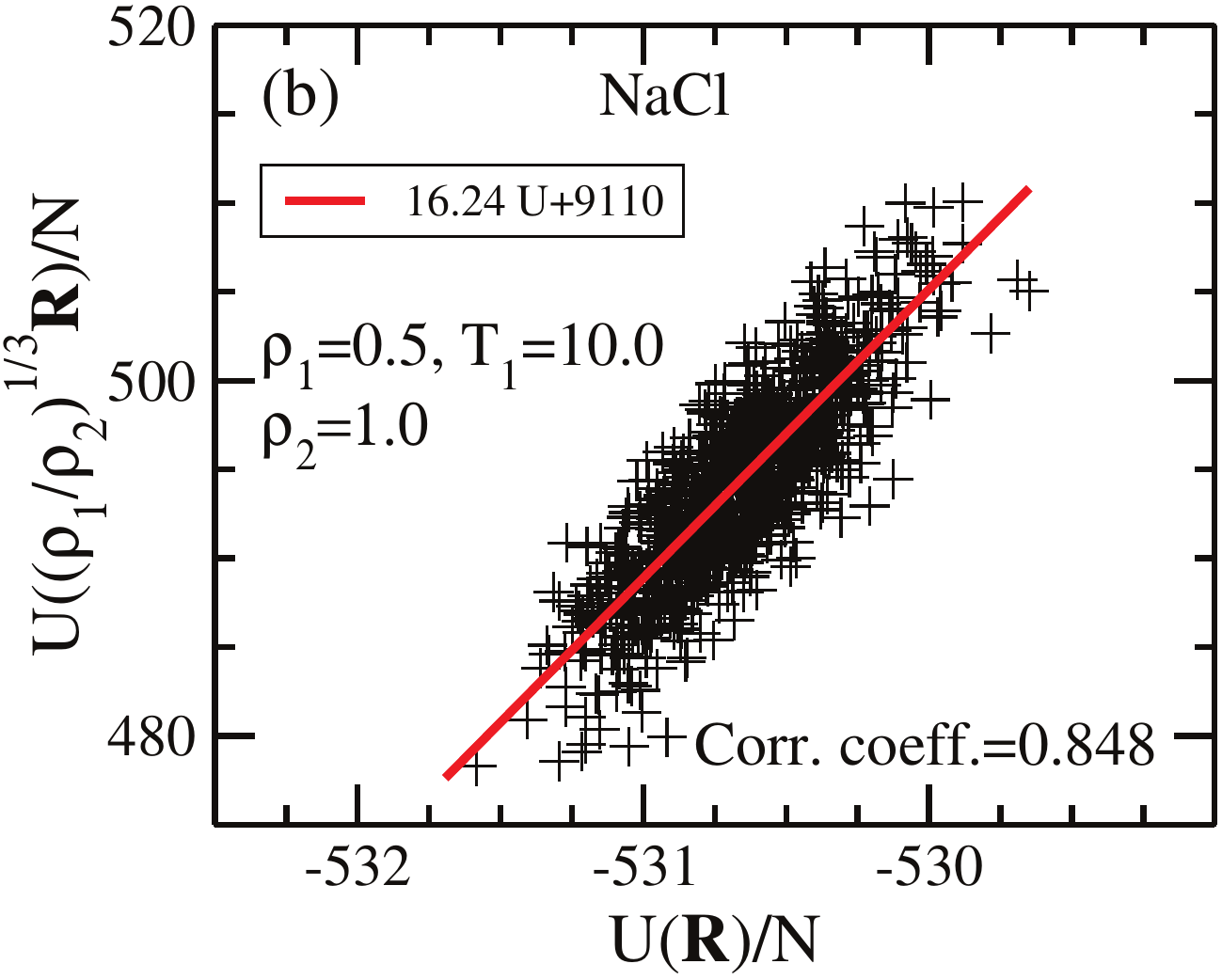}
\includegraphics[width=.3\textwidth]{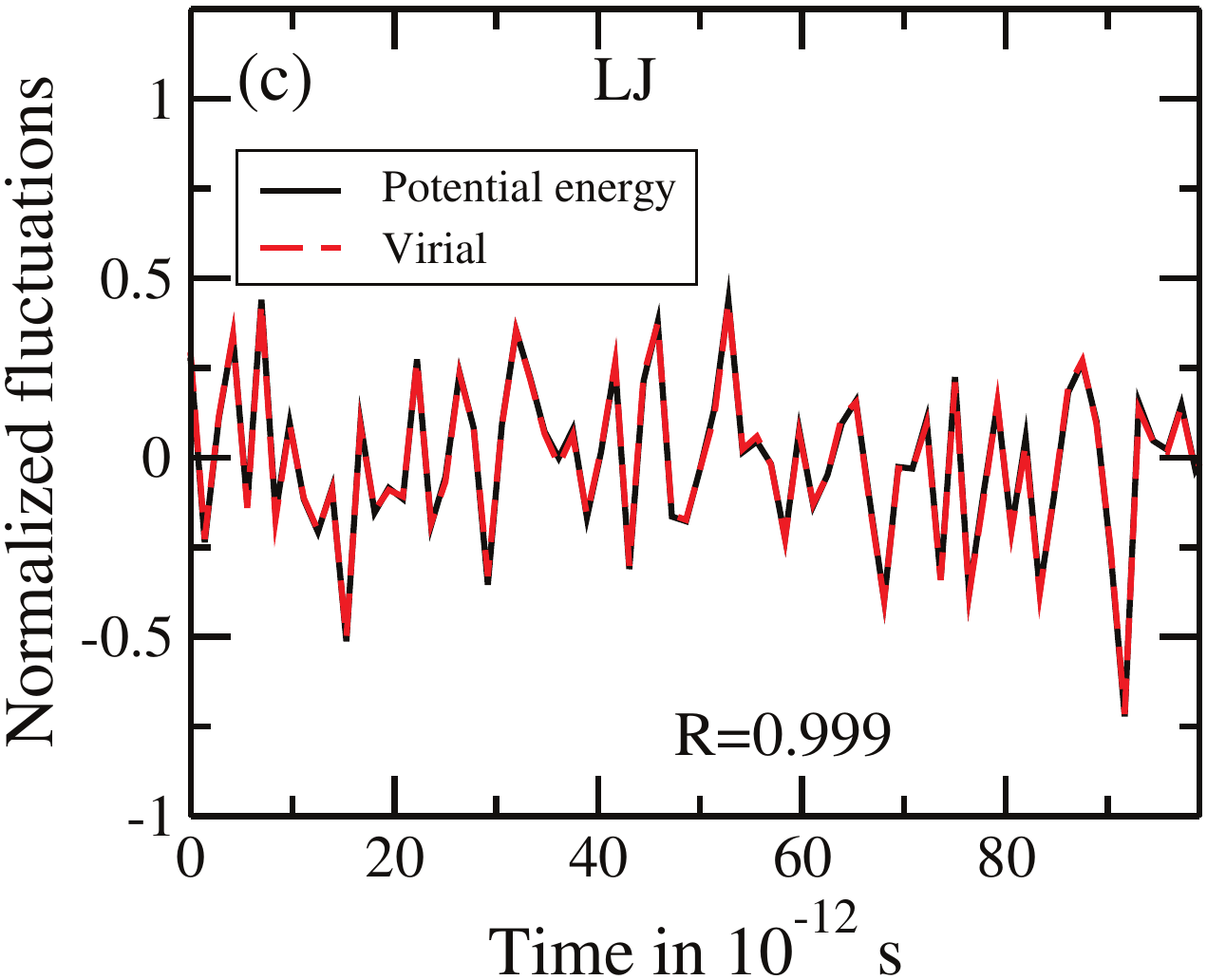}
\includegraphics[width=.3\textwidth]{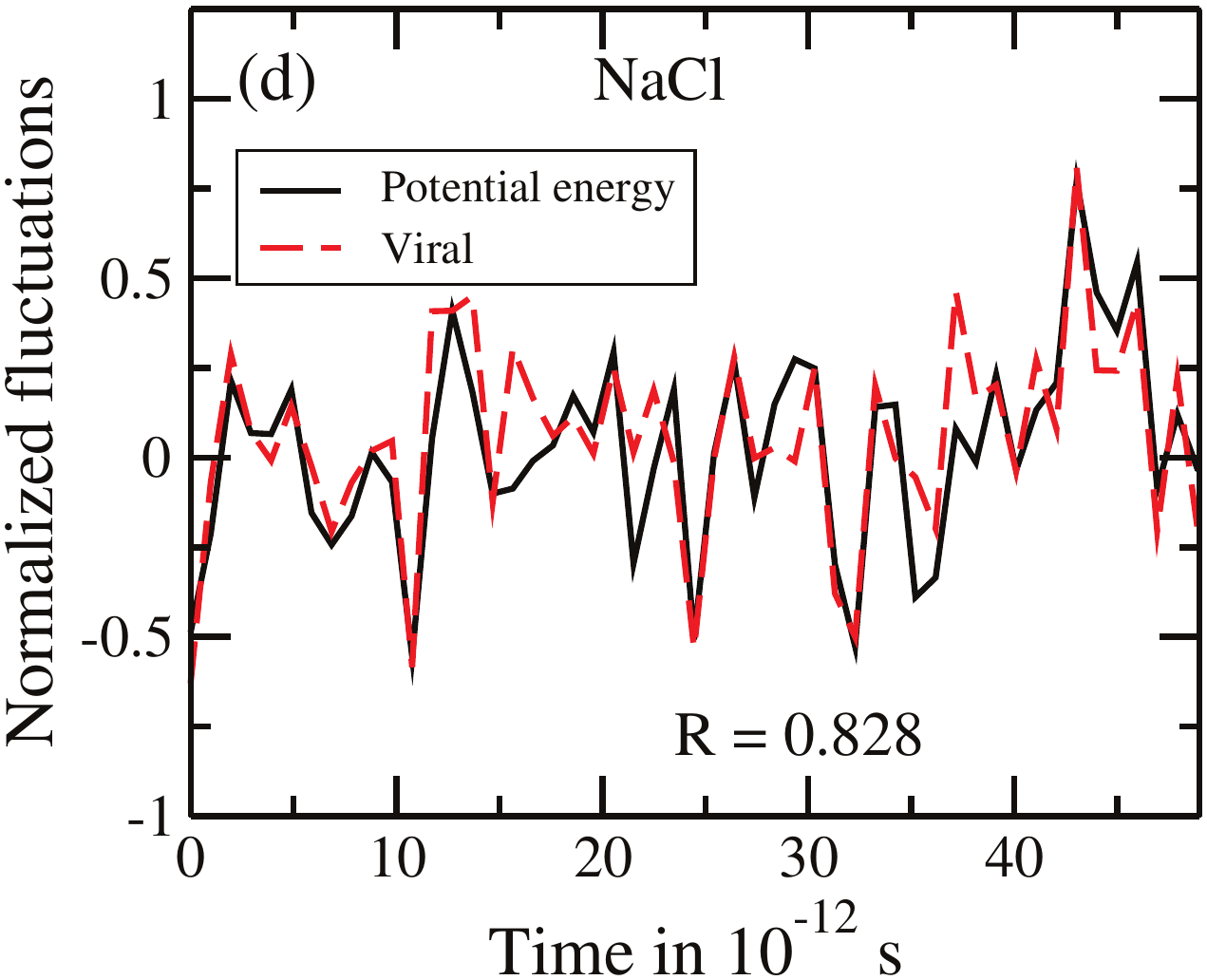}
\caption{(a) and (b) Scatter plots of the potential energies of scaled configurations.
(a) Results from $NVT$ simulations of a single-component Lennard-Jones (LJ) $10\times 10\times 10$ face-centered cubic (FCC) crystal at density $\rho_1=1.2$ and temperature $T_1=1.0$. The figure shows a scatter plot of the potential energy per particle when each configuration is scaled uniformly to density $\rho_2=2.4$, versus the potential energy of the original configuration denoted by $\bR$ at density $\rho_1$. Note that upon the compression the potential energies change from small negative to large positive values. The blue dashed line of slope $16=(2^{-1/3})^{-12}$ gives the prediction of the repulsive $r^{-12}$ part of the LJ potential (plus a constant); this does not fit the data that have slope 21.14. 
(b) Similar scatter plot for simulations of a model NaCl crystal \cite{smi94} in which density is also doubled.
(c) Plot of virial and potential energy as functions of time in Argon units for the LJ crystal in equilibrium at the state point $(\rho_1,T_1)$, demonstrating very strong correlations;
(d) similar plot for the NaCl model, showing much weaker correlations. The correlation coefficient $R$ given in figures (c) and (d) is defined in Eq. (\ref{R}).}
\label{DIC_lj}
\end{figure}

The first step in the investigation of the LJ crystal is to identify the isomorphs in the thermodynamic phase diagram; these are identical to the configurational adiabats \cite{IV} (see below). Consider a FCC crystal in thermal equilibrium at density $\rho_1=1.2\sigma^{-3}$ and temperature $T_1=1.0\,\varepsilon/k_B$, which in standard LJ units is written as follows: $\rho_1=1.2$ and $T_1=1.0$. The idea is now that for an equilibrium simulation at this state point, each configuration is scaled uniformly to a different density. In Fig. \ref{DIC_lj}(a) the scaling is to the double density, $\rho_2=2.4$. Each configuration is characterized by the $3N$-dimensional configuration vector $\bR \equiv (\br_1,...,\br_N)$, and if the original and the scaled configurations are denoted by $\bR_1$ and $\bR_2$, respectively, the two configurations $\bR_1$ and $\bR_2$ have the same reduced coordinates and $\bR_2=2^{-1/3}\bR_1$.

Figure \ref{DIC_lj}(a) shows a scatter plot of the potential energies of pairs of scaled versus original configuration generated in an equilibrium simulation at $(\rho_1,T_1)$. The potential energies of original and scaled configurations correlate strongly. That this is not a trivial effect of the crystal structure and, for instance, a crystal's approximate harmonic description, is clear from Fig. \ref{DIC_lj}(b) giving similar data for the NaCl crystal model \cite{smi94}. The blue dashed line in Fig. \ref{DIC_lj}(a) is the prediction of the repulsive $r^{-12}$ part of the LJ pair potential (displaced vertically by a large amount to fit the data); clearly this term alone cannot explain the strong correlation observed \cite{II}. Strong correlations of the potential energies of scaled configurations have been reported for several liquids, including various LJ-type and molecular model liquids \cite{ing12,ing12b}, and in fact even for the 10-bead rigid-bond, flexible LJ-chain model \cite{vel14}. 

The linear correlation of Fig. \ref{DIC_lj}(a) implies that a temperature $T_2$ exists such that the following applies. Whenever two configurations $\bR_1$ and $\bR_2$ of the densities $\rho_1$ and $\rho_2$, respectively, have the same reduced coordinates, i.e., $\rho_1^{1/3}\bR_1=\rho_2^{1/3}\bR_2$, one has 
$U(\bR_2)\cong(T_2/T_1)U(\bR_1)+{\rm Const}.$ This implies the isomorph condition \cite{IV}

\be\label{iso_def}
\exp\left({-\frac{U(\bR_1)}{k_BT_1}}\right)
\,\cong\,C_{12}\,\exp\left({-\frac{U(\bR_2)}{k_BT_2}}\right)\,.
\ee
The temperature $T_2$ is determined from the best fit line slope of Fig. \ref{DIC_lj}(a), which is $21.14$, as follows: $T_2=21.14\cdot T_1=21.14$.  

Equation (\ref{iso_def}) implies (almost) identical canonical-ensemble probabilities of pairs of scaled configurations belonging to the state points $(\rho_1,T_1)$ and $(\rho_2,T_2)$. As a consequence, the configurational (``excess'') entropy of the two state points are (almost) identical \cite{IV}. In fact, Eq. (\ref{iso_def}) implies that the structure and dynamics of two isomorphic state points are (almost) identical when given in reduced units \cite{IV}. 

Two state points are termed {\it isomorphic} if they obey Eq. (\ref{iso_def}) for their physically relevant configurations \cite{IV}. In this way a mathematical equivalence relation is defined in the thermodynamic phase diagram; its equivalence classes are continuous curves termed isomorphs. The above method of identifying isomorphic state points via a scatter plot of scaled potential energies is referred to as the ``direct isomorph check'' \cite{IV}.

An isomorph is a configurational adiabat because the excess entropy is an isomorph invariant. The opposite does not apply in general, however -- all systems have configurational adiabats, but only some systems have isomorphs. The microscopic virial is defined by $W(\bR)\equiv -1/3\bR\cdot\nabla U(\bR)$ \cite{tildesley}. It has been shown that a system has isomorphs if and only if it has strong virial potential-energy correlations for its  thermal equilibrium constant-volume fluctuations (Appendix A of Ref. \onlinecite{IV}). A pragmatic definition of a Roskilde system is $R>0.9$ in which $R$ is the virial potential-energy (Pearson) correlation coefficient defined \cite{I} by (where $\Delta W$ is the deviation from the average virial, $\Delta U$ the same for $U$, and the sharp brackets denote canonical averages)

\be\label{R}
R\equiv\frac{\langle\Delta U\Delta W\rangle}{\sqrt{\langle(\Delta U)^2\rangle\langle(\Delta W)^2\rangle}}\,.
\ee

It follows from Euler's theorem that systems with a homogeneous potential-energy function, i.e., those for which $U(\lambda\bR)=\lambda^{-n}U(\bR)$ for some exponent $n$, are the only ones with 100\% virial potential-energy correlations and thus exact isomorphs. This means that any realistic system's isomorph invariants are only approximate. Figures \ref{DIC_lj}(c) and (d) show the normalized virial and potential-energy equilibrium fluctuations as functions of time for the LJ and NaCl crystals. Only the LJ crystal exhibits strong virial potential-energy correlations, so only for this system is the isomorph theory expected to work \cite{IV}.

\begin{figure}
\centering
\includegraphics[width=.3\textwidth]{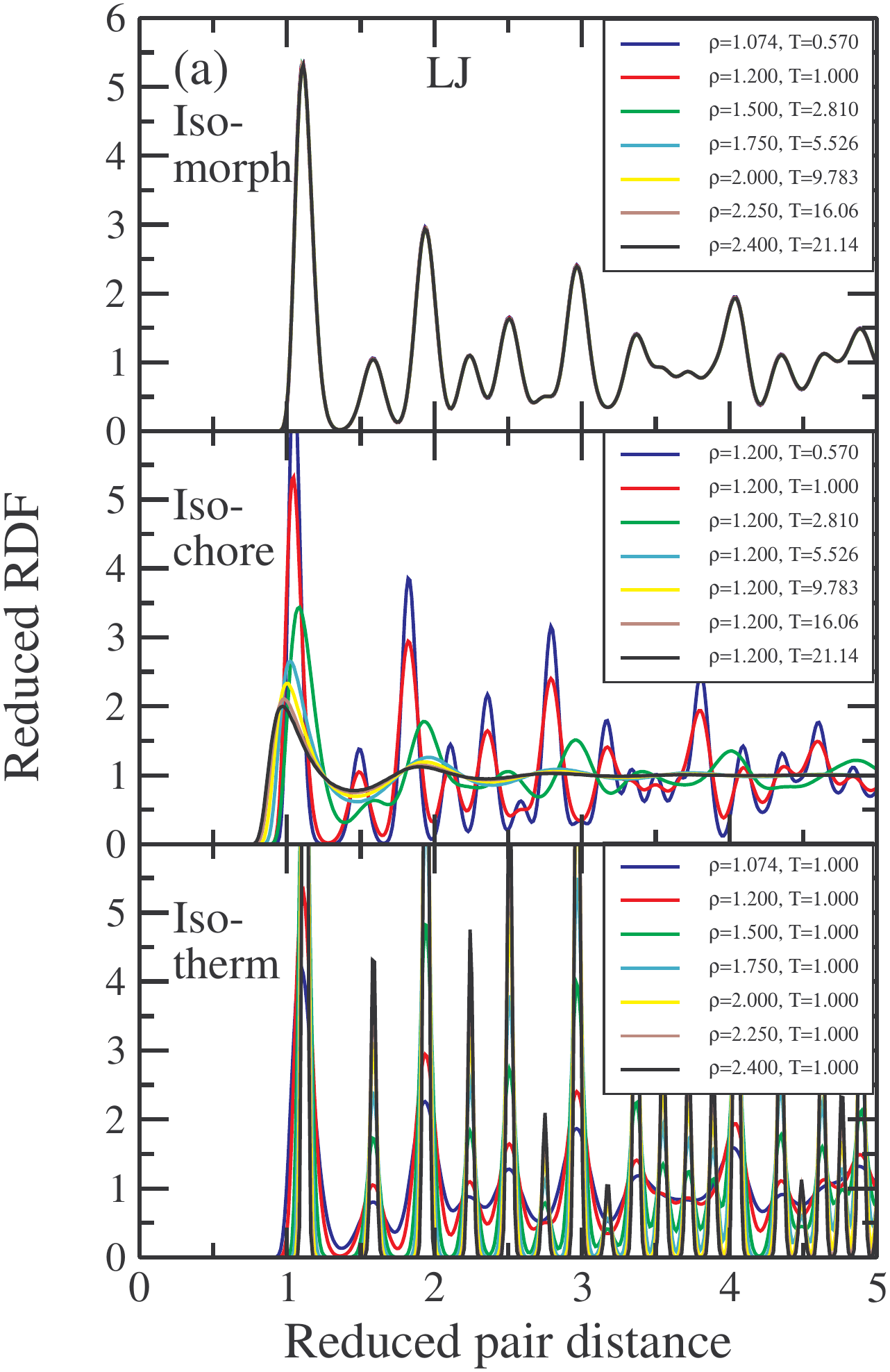}
\includegraphics[width=.3\textwidth]{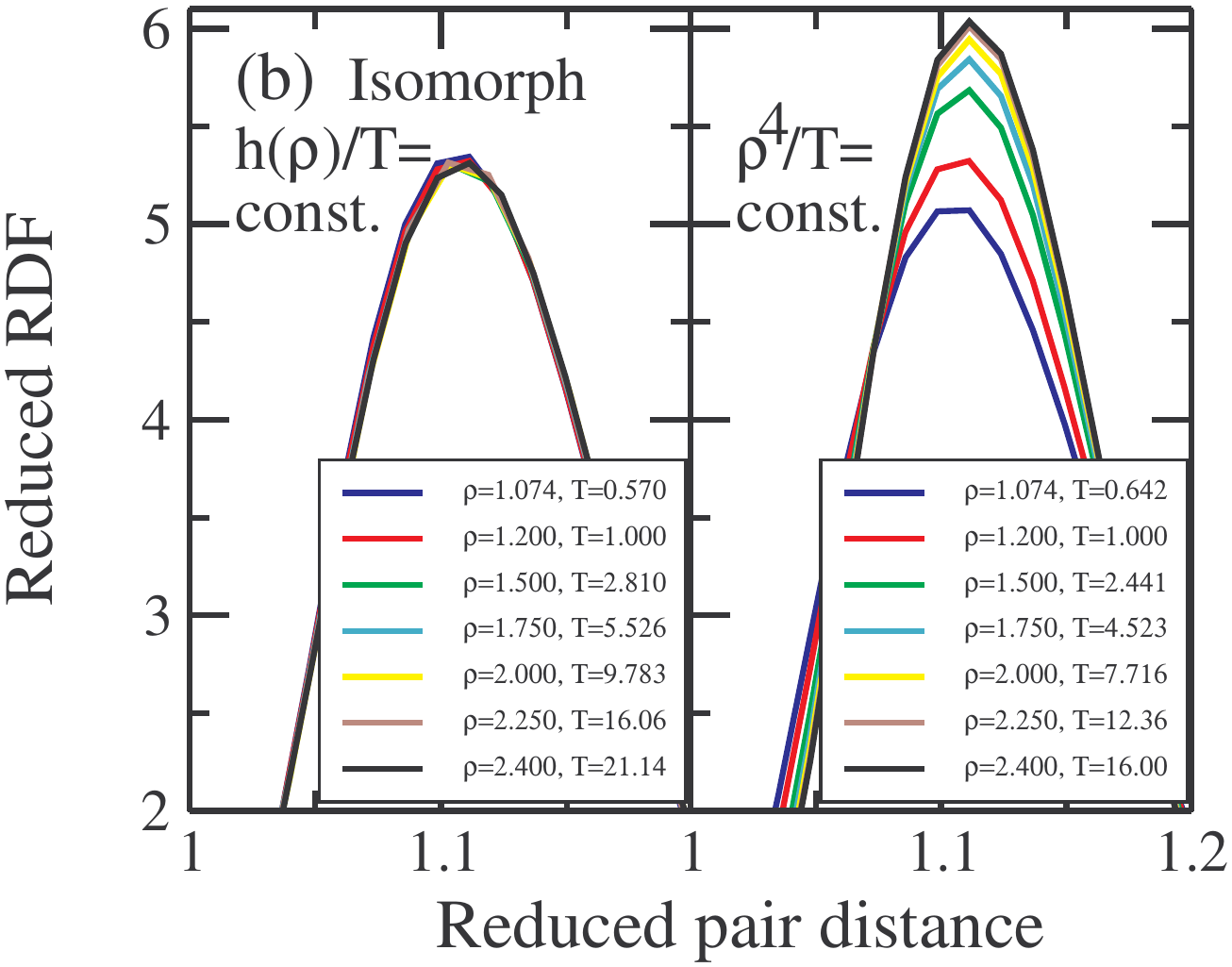}
\caption{Radial distribution functions (RDF) of the LJ crystal as functions of the reduced pair distance $\tilde r$ of Eq. (\ref{rred}) along various lines in the thermodynamic phase diagram.
(a, upper subfigure) RDFs calculated for state points along an isomorph, involving more than a factor of two density change. The data collapse demonstrates structural invariance. For comparison, the bottom two figures of (a) give the RDFs from state points along an isochore and an isotherm, respectively, for the same temperature / density variation.
(b) A zoom-in on the first peak of the RDF for isomorph (left) and $r^{-12}$ inverse-power-law scaling implying invariance along the line of constant $\rho^4/T$ (right), demonstrating that isomorph invariance is not merely a trivial consequences of the repulsive $r^{-12}$ term of the LJ pair potential.}
\label{RDF}
\end{figure}

It was recently shown that the properties of Roskilde systems, including the existence of isomorphs, is a consequence of these systems' ``hidden scale invariance'' by which is meant the following: two functions of density exist, $h(\rho)$ and $g(\rho)$, such that for all physically relevant configurations one has \cite{dyr13a,dyr14}

\be\label{prop1m}
U(\bR)\cong h(\rho)\tU(\rho^{1/3}\bR)+g(\rho)\,.
\ee
Here $\tU$ is a dimensionless function of $\rho^{1/3}\bR$, the reduced configuration vector. For a system of particles interacting via the pair potential $v(r)=\sum_n \varepsilon_n (r/\sigma)^{-n}$, the functions $h(\rho)$ and $g(\rho)$ are both given by expressions of the form $\sum_n C_n \rho^{n/3}$ in which each $\rho^{n/3}$ term corresponds to a $(r/\sigma)^{-n}$ term in $v(r)$ \cite{dyr13a}.

The physical content of Eq. (\ref{prop1m}) is that a density change to a good approximation results simply in a linear affine scaling of the potential-energy surface. Since the addition of an overall density-dependent constant to the potential energy does not affect structure and dynamics -- though it does, of course, change the free energy and the pressure -- the affine scaling of Eq. (\ref{prop1m}) can be compensated by adjusting the temperature in proportion to $h(\rho)$, leading to identical canonical probabilities of configurations that scale uniformly into one another \cite{ing12a}. This is what happens along an isomorph, resulting in reduced-unit invariance of structure and dynamics \cite{IV,dyr14}. Thus Eq. (\ref{prop1m}) implies the existence of isomorphs. This equation can also be shown to imply the following thermodynamic separation identity $k_BT=f(\sex)h(\rho)$ in which $\sex$ is the excess entropy per particle \cite{ing12a}. Since isomorphs are configurational adiabats, this identity implies that the isomorphs are given by the equation $h(\rho)/T={\rm Const.}$ We finally note \cite{nag11,ing12a} that the separation identity $k_BT=f(\sex)h(\rho)$ is mathematically equivalent to the configuration-space version of the noted Gr{\"u}neisen equation of state, according to which pressure is a linear function of energy with constants that only depend on density. This is the standard equation of state for solids under high pressure \cite{nag11}.

\begin{figure}
\centering
\includegraphics[width=.3\textwidth]{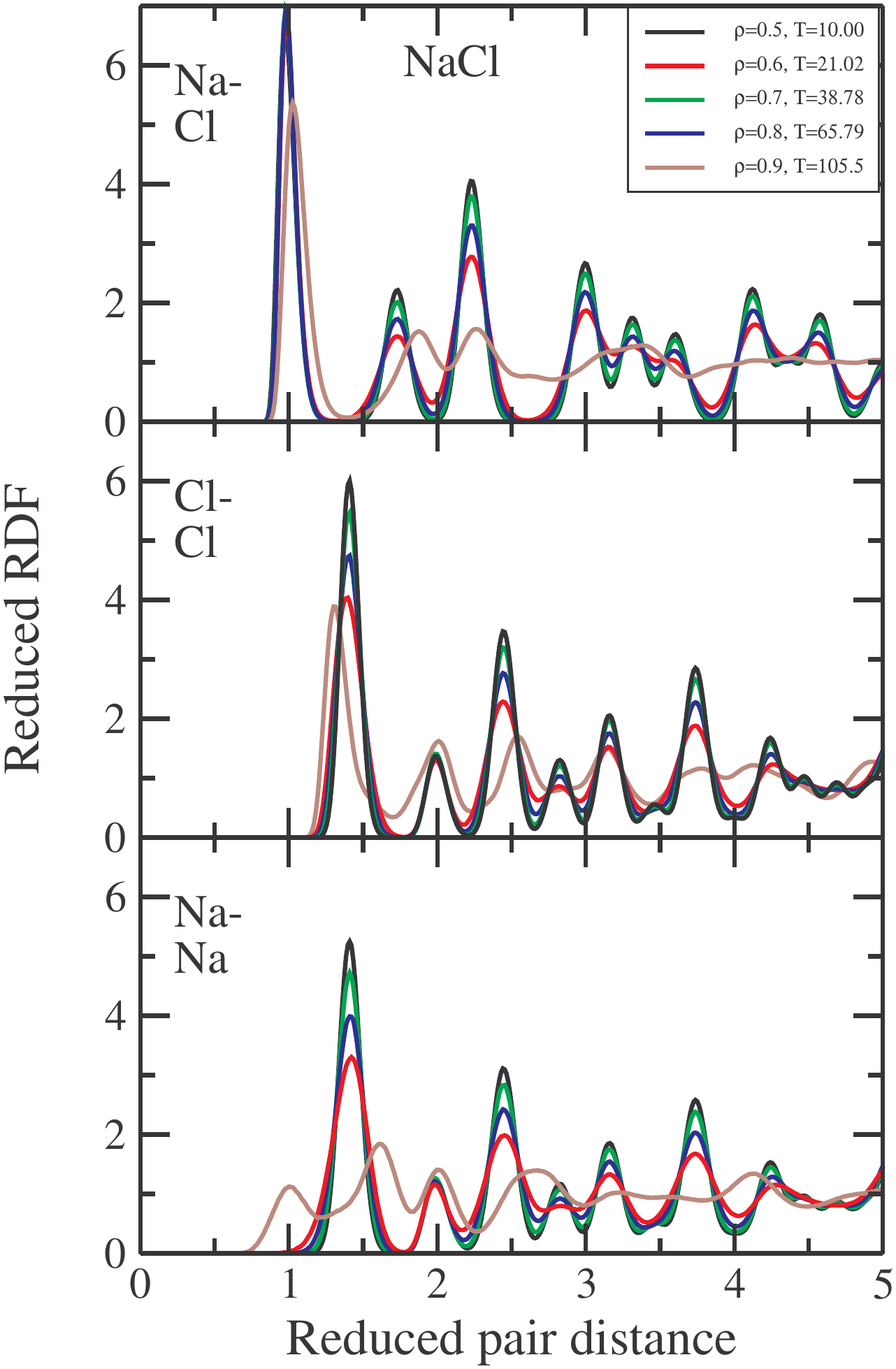}
\caption{RDFs of the NaCl crystal for prospective isomorphic state points identified by the direct isomorph check method (Fig. \ref{DIC_lj}(b)). The figure shows the sodium-chloride, chloride-chloride, and sodium-sodium partial RDFs. This crystal does not have isomorphs (at the highest-density state point simulated the systems in fact has melted).}
\label{RDF_NaCl}
\end{figure}

\section{Isomorph invariance of structure and dynamics: Comparing the LJ and sodium chloride crystals}\label{structure}

This section presents simulations of some properties predicted to be invariant along an isomorph. The purpose is to validate the existence of isomorphs for the LJ crystal. The LJ simulations are contrasted to simulations of the NaCl crystal for which, based on Fig. \ref{DIC_lj}, one does not expect isomorphs. Prospective isomorphic state points were identified using the direct isomorph-check method described above for simulations at the following reference state points: $(\rho,T)=(1.2,1)$ for the LJ crystal and $(\rho,T)=(0.5,10)$ for the NaCl crystal.

Figure \ref{RDF}(a) shows the radial distribution function (RDF) of the LJ crystal at different state points as a function of the reduced pair distance $\tilde r$ defined by 

\be\label{rred}
\tilde r \equiv\rho^{1/3}r\,.
\ee
The upper subfigure shows RDFs of the LJ crystal along an isomorph, predicted to be invariant. We see that this applies to very good approximation for more than a factor of two density change along the isomorph, which is represented by seven state points. For comparison, the two lower subfigures give the reduced-unit RDFs for the same temperature/density variations keeping density and temperature constant, respectively. 

Most of the state points studied correspond to highly compressed solids, so one might guess that the observed structural invariance derives trivially from the repulsive $r^{-12}$ term dominating the LJ potential. If this were the case, however, state points related by $\rho^4/T={\rm Const.}$ should have the same reduced-unit RDF \cite{III}. Figure \ref{RDF}(b) shows a blow up of the first peak of the reduced-unit RDFs along the curve defined by $\rho^4/T={\rm Const.}$ for the same density variation (right subfigure, same reference state point). There is poor collapse compared to the left subfigure, which is a blow up of the Fig. \ref{RDF}(a) isomorph data. Thus the isomorph collapse is not a trivial consequence of the LJ potential's repulsive $r^{-12}$ term, which was clear already from Fig. \ref{DIC_lj}(a) in which the scatter plot does not have the slope 16 predicted from the $r^{-12}$ term of the LJ potential.

Isomorph invariance of structure is not a general property of crystals. This is evident from Fig. \ref{RDF_NaCl}, which shows RDFs for prospective isomorphic state points of the NaCl crystal determined by the direct-isomorph-check procedure. No structural invariance is observed. Thus isomorph invariance is not a trivial consequence of  the harmonic approximation that generally describes crystals well.

\begin{figure}
\centering
\includegraphics[width=.3\textwidth]{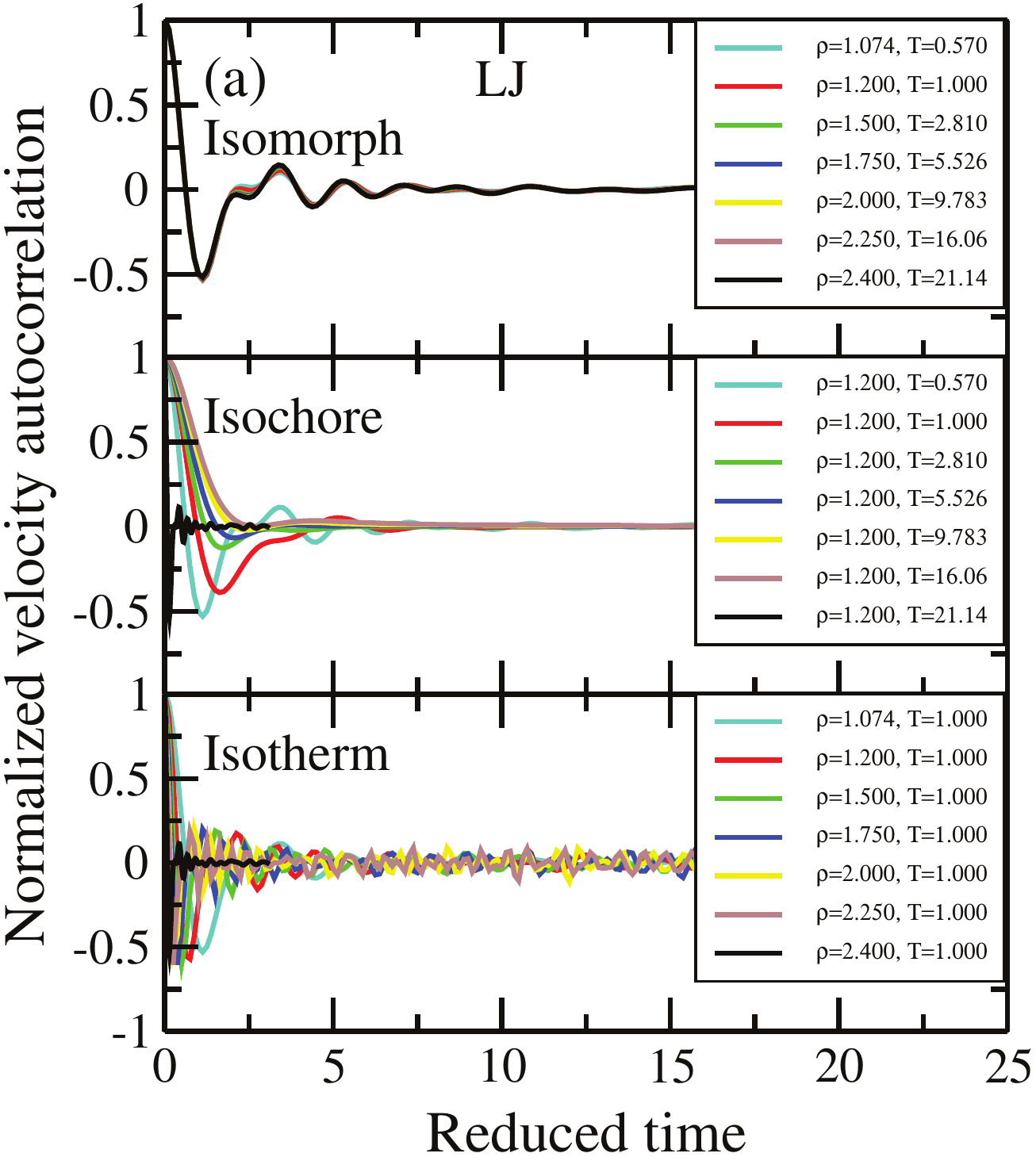}
\includegraphics[width=.3\textwidth]{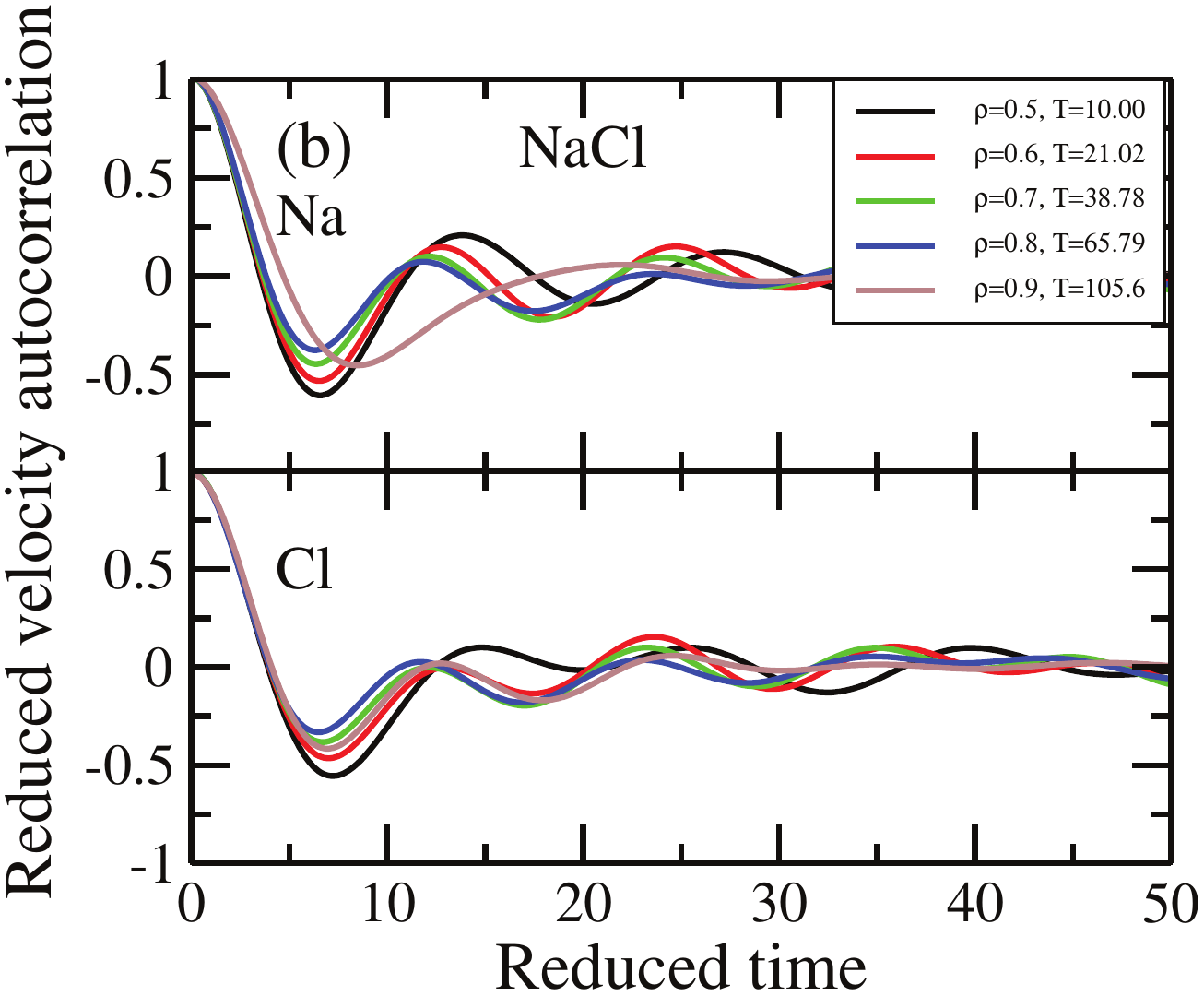}
\caption{
(a) Normalized velocity autocorrelation functions of the LJ crystal along an isomorph, an isochore, and an isotherm (same state points as in Fig. \ref{RDF}). 
(b) Normalized velocity autocorrelation functions for the NaCl crystal along a prospective isomorph, showing no data collapse (same state points as in Fig. \ref{RDF_NaCl}).}
\label{VAF}
\end{figure}

Since the potential-energy surfaces of isomorphic state points are identical except for a linear, affine scaling (compare the hidden-scale-invariance identity Eq. (\ref{prop1m})), not just the structure, but also the dynamics is predicted to be invariant along an isomorph when given in reduced units \cite{IV,dyr14}. For the LJ crystal we checked this by calculating the single-particle velocity autocorrelation function. Figure \ref{VAF}(a) shows the results for the state points of Fig. \ref{RDF}. Good collapse is found for the isomorphic state points (upper subfigure), although the collapse is not quite as good as for the RDFs. The NaCl crystal shows no data collapse  (Fig. \ref{VAF}(b)). 

Our simulations show that the phonon dynamics of the LJ crystal is isomorph invariant. Perfect crystals like the one simulated have no slow dynamics. In order to test the generality of the isomorph invariance of the dynamics, we investigated also the much slower dynamics of atoms jumping in a defective LJ crystal obtained by removing some particles. The vacancies thus introduced occasionally jump to new lattice positions, a process making atomic diffusion possible that, because it is thermally activated, becomes slow at low temperatures and/or high densities. 

\begin{figure}
\centering
\includegraphics[width=0.3\textwidth]{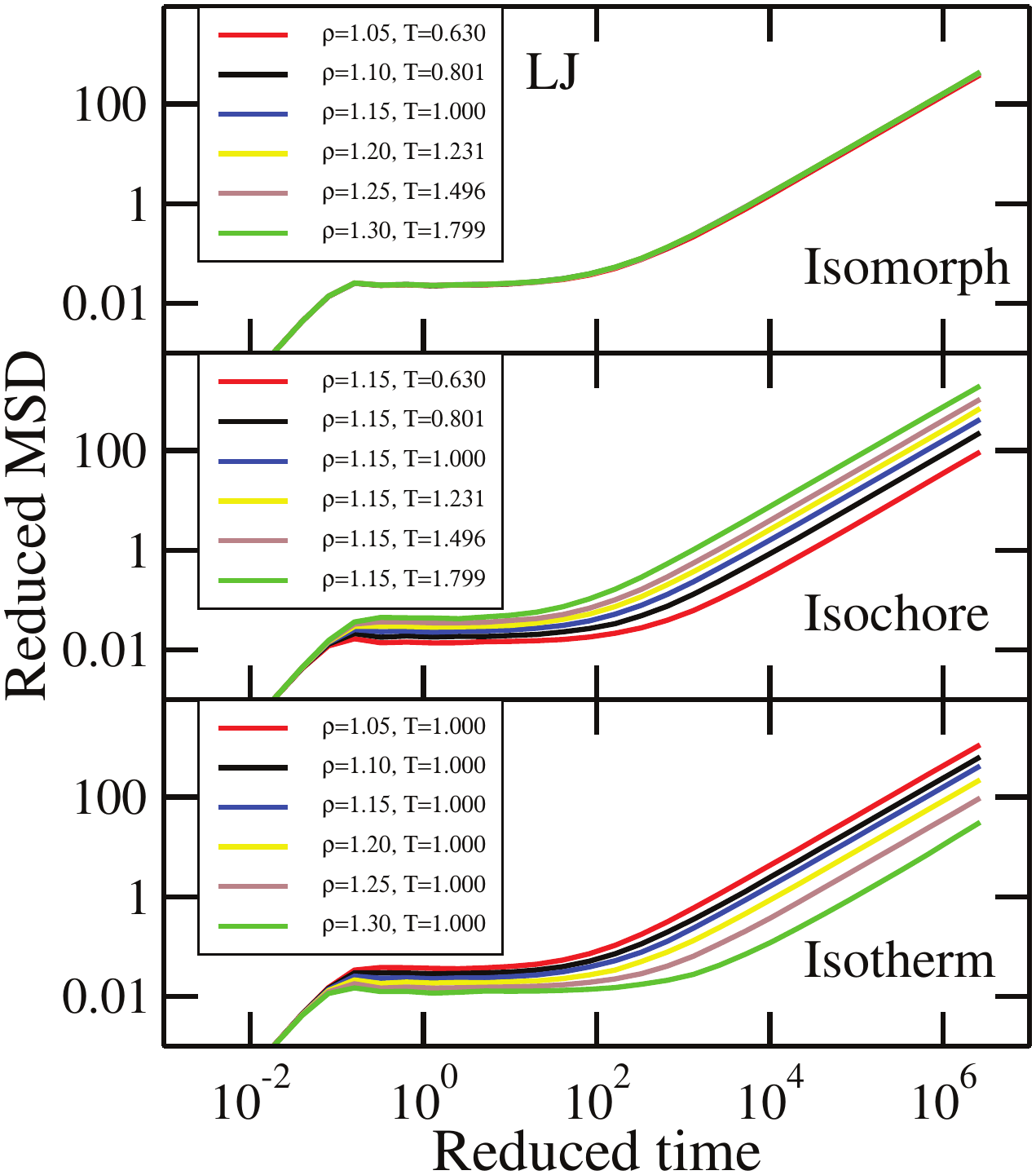}
\caption{Reduced-unit mean-square displacement (MSD) of the particles of an $8\times 8\times 8$ FCC LJ crystal from which eight particles were removed, making vacancy-jump dynamics possible. The top figure shows the MSD as a function of reduced time for state points along an isomorph identified by the direct-isomorph-check method. The two bottom figures show MSDs calculated for state points along the isochore and isotherm with the same temperature/density variation. Vacancy jump dynamics depends strongly on temperature and density, but along an isomorph these two effects compensate each other to a  good approximation.}
\label{VAC}
\end{figure}

Eight particles were removed from a LJ crystal, thus introducing eight vacancies. The effect of vacancy jumps was monitored by evaluating the mean-square displacement (MSD) of the defective crystal's particles as a function time. The results are shown in Fig. \ref{VAC} for state points along an isomorph, an isochore, and an isotherm. The isomorph was generated by the direct isomorph check method applied to the defective crystal starting at $(\rho,T)=(1.15,1)$, leading to an isomorph that is marginally different from one of the perfect LJ crystal. Figure \ref{VAC} demonstrates that the regime of slow atomic jump dynamics is also isomorph invariant to a good approximation. This is nontrivial, even in view of the isomorph invariance of structure and fast dynamics of perfect crystals demonstrated in Figs. \ref{RDF} and \ref{VAF}, because the excitation states of vacancy jumps -- the barriers to be overcome -- have a small canonical probability and contribute little to the direct-isomorph-check plots used to identify isomorphic state points (Fig. \ref{DIC_lj}).

\section{Isomorph jumps of the LJ crystal}\label{jump}

Because the canonical-ensemble probabilities of scaled configurations of isomorphic state points are identical, a jump between isomorphic state points is predicted to bring the system instantaneously to equilibrium \cite{IV}. Instantaneous equilibration after isomorph jumps has been demonstrated in simulations of supercooled, highly viscous liquids \cite{IV,ing12b}, as well as for the flexible LJ-chain model \cite{vel14}. Below we test whether this applies also for ``isomorph jumps'' of perfect LJ crystals, for which the dynamics takes place on the much faster phonon time scale of order picoseconds (in Argon units). 

\begin{figure}
\centering
\includegraphics[width=.3\textwidth]{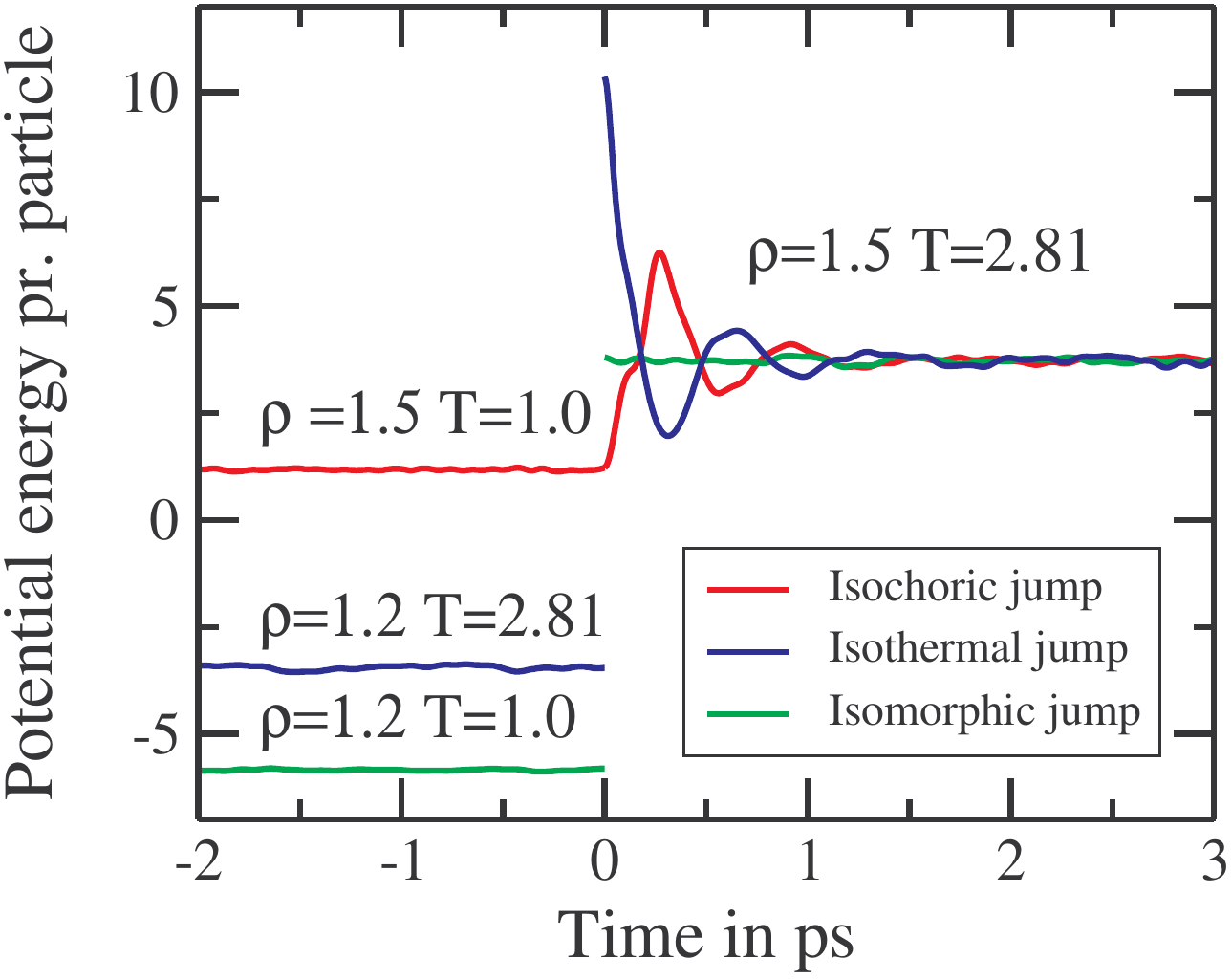}
\caption{Potential energy per particle after jumps from three different state points of the LJ crystal to the state point $(\rho,T)=(1.5, 2.813)$. Only the jump from a state point isomorphic to the final state point (green curve) leads to instantaneous equilibration. Time is given in Argon units.}
\label{JUMP}
\end{figure}

The procedure used for studying a jump in thermodynamic phase space is the following. First one equilibrates the system at one state point. Then one changes the density by scaling all coordinates uniformly, scaling all velocities to the new temperature, and adjusting the temperature of the thermostat to the new value. Finally, one observes whether or not the system relaxes at the new state point by monitoring the time development of a suitable quantity, {\it in casu} the potential energy. Figure \ref{JUMP} shows the potential energy per particle after such jumps from three different state points to the same state point. The initial state points were selected to give an isochoric (red), an isothermal (blue), and an isomorphic (green) jump to the final state point. Only the latter shows instantaneous equilibration, i.e., no change of the potential energy after the jump.

\section{Results for five other model crystals}\label{other}

A number of questions arise from our findings for the LJ and NaCl crystals. To answer some of these we present below simulation results for more systems, three atomic and two molecular crystals.

One question is whether complex crystals may also have isomorphs or this property is limited to simple close-packed crystal structures like the FCC lattice. Liquid-state simulations show that the property of strong virial potential-energy correlations -- implying isomorphs \cite{IV} -- is ``local'' in the sense that for Roskilde systems all interactions beyond the first coordination shell are unimportant and may be ignored \cite{ing12}. Based on this one would not expect the crystal structure to be  important. In order to illuminate this issue we made use of the recently discovered fact \cite{ped10b} that the Wahnstr{\"o}m binary LJ system \cite{wah91} crystallizes into the intricate ${\rm MgZn_2}$ Laves phase, in which the unit cell contains no less than twelve atoms. Laves phases are observed in some binary metal systems \cite{graef} and, e.g., binary hard-sphere mixtures of certain size ratios crystallize into Laves phases. Laves phases of metallic elements have a number of intriguing properties, for instance they are not plastically deformable at room temperature \cite{wiki}. 

The ${\rm MgZn_2}$ Laves phase of the Wahnstr{\"o}m binary LJ model is characterized as follows \cite{ped10b}. The smaller LJ particles are placed in one of two different distorted icosahedral polyhedra, both made up of six small and six large LJ particles. The large particles sit in a 16-vertex coordination polyhedron comprised of twelve small and four large particles. The latter form a hexagonal diamond network where each large-particle neighbor pair shares six small particle neighbors.

\begin{figure}
\centering
\includegraphics[width=.3\textwidth]{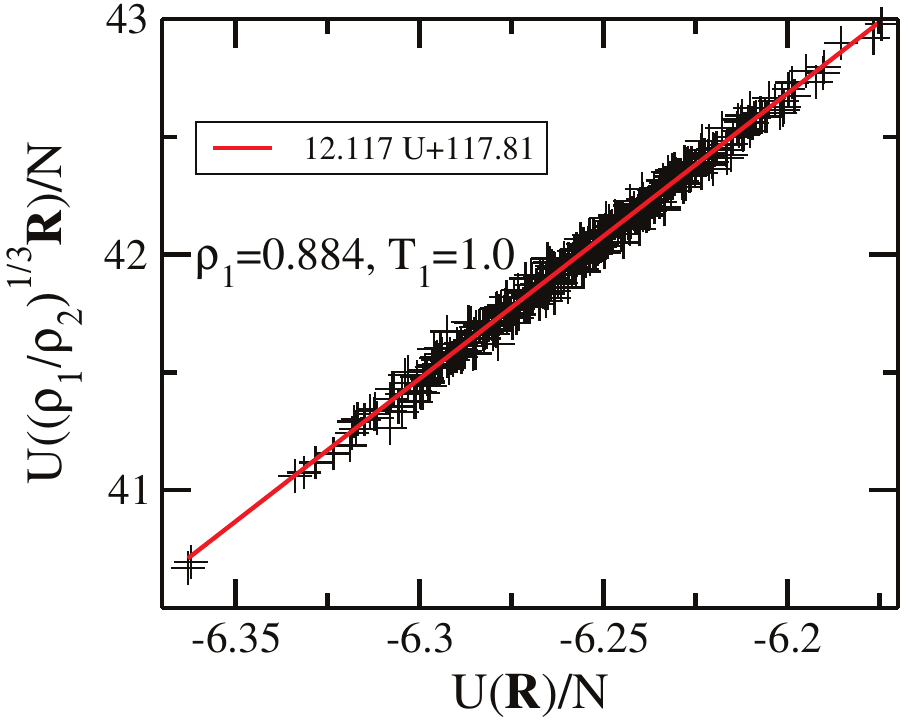}
\includegraphics[width=.3\textwidth]{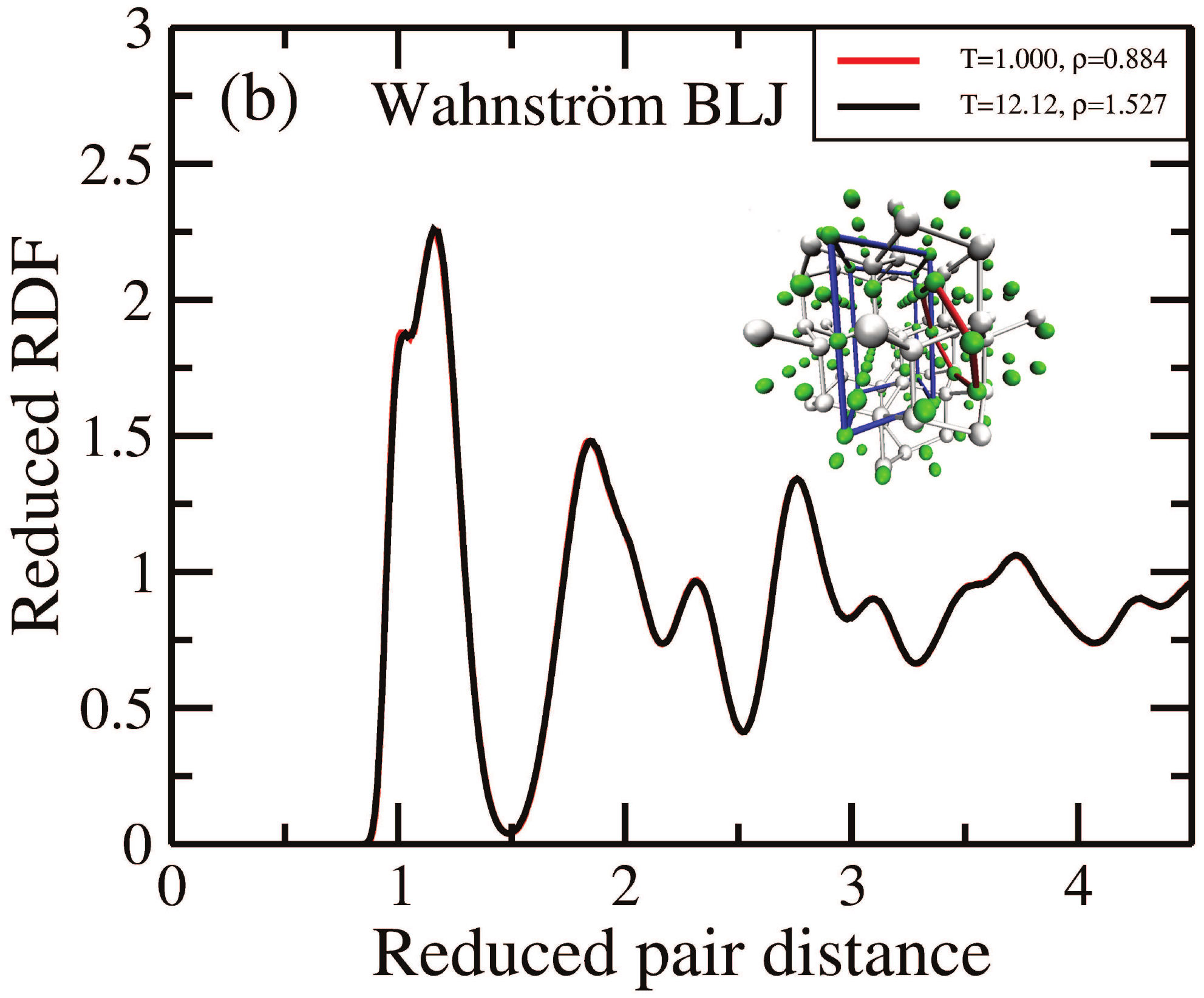}
\caption{(a) Direct isomorph check for the Wahnstr{\"o}m binary LJ crystal, which is a ${\rm MgZn_2}$ Laves phase structure with a unit cell of twelve atoms. 
(b) All-particle RDF for the two isomorphic state points identified on the basis of the direct isomorph check in (a). The inset shows the crystal structure \cite{ped10b}.}
\label{BiLJ}
\end{figure}

Figure \ref{BiLJ} shows results from simulations of this system. For details of the LJ parameters we refer to Ref. \onlinecite{wah91}; our only modification was to consider a perfect crystal of 2/3 small and 1/3 large particles whereas the original Wahnstr{\"o}m paper considered a 50-50 composition in the supercooled liquid state. Figure \ref{BiLJ}(a) shows direct-isomorph-check results for a 36\% density increase. Figure \ref{BiLJ}(b) shows the RDF for all particles (small and large) at two isomorphic state points with the temperature at the high-density state point calculated from the direct isomorph check of Fig. \ref{BiLJ}(a) ($T_2=12.12$). We see that even complex crystal structures can have isomorphs. We moreover conclude that the reason the NaCl crystal does not have isomorphs is not simply that it is a two-component system.

\begin{figure}
\centering
\includegraphics[width=.3\textwidth]{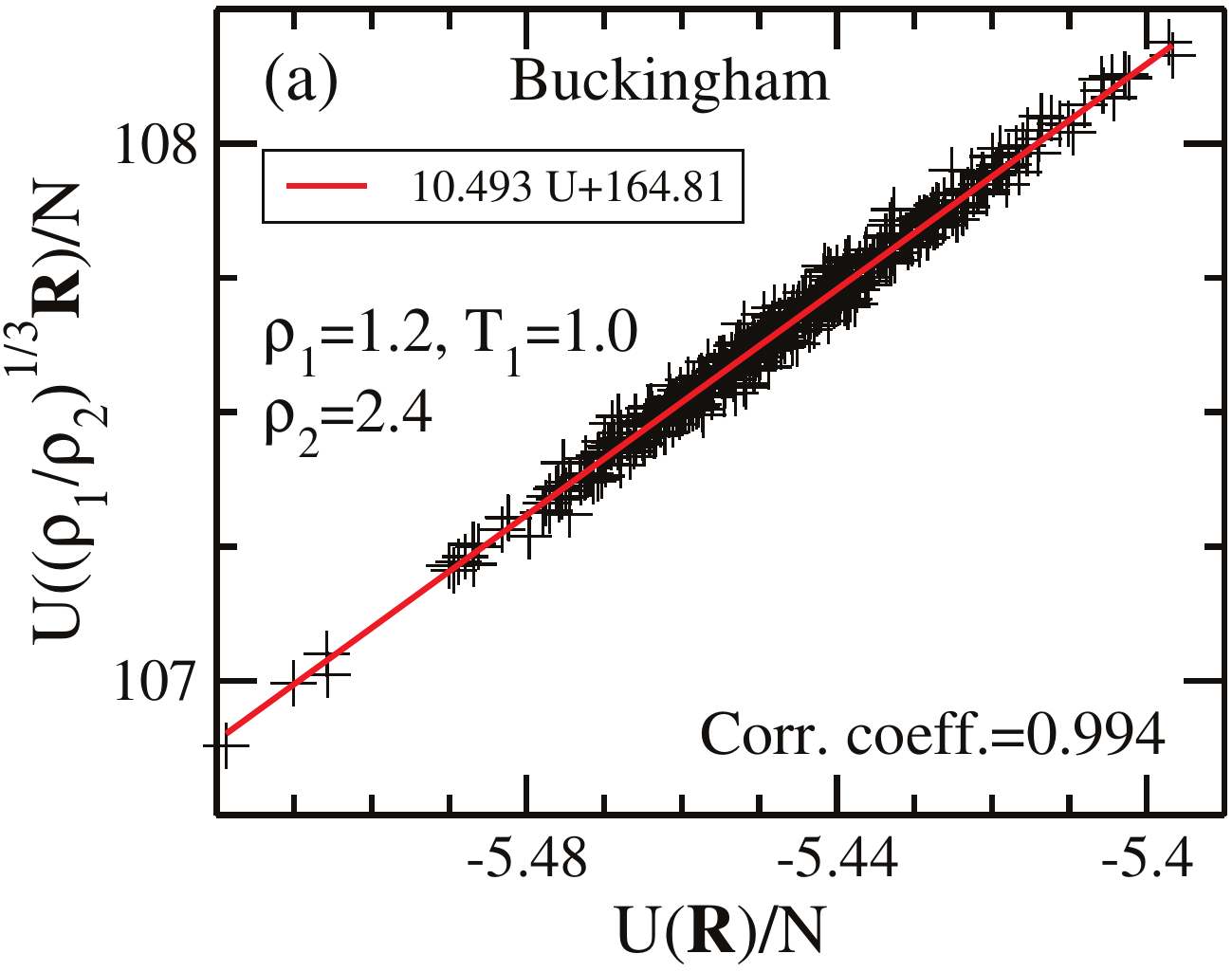}
\includegraphics[width=.3\textwidth]{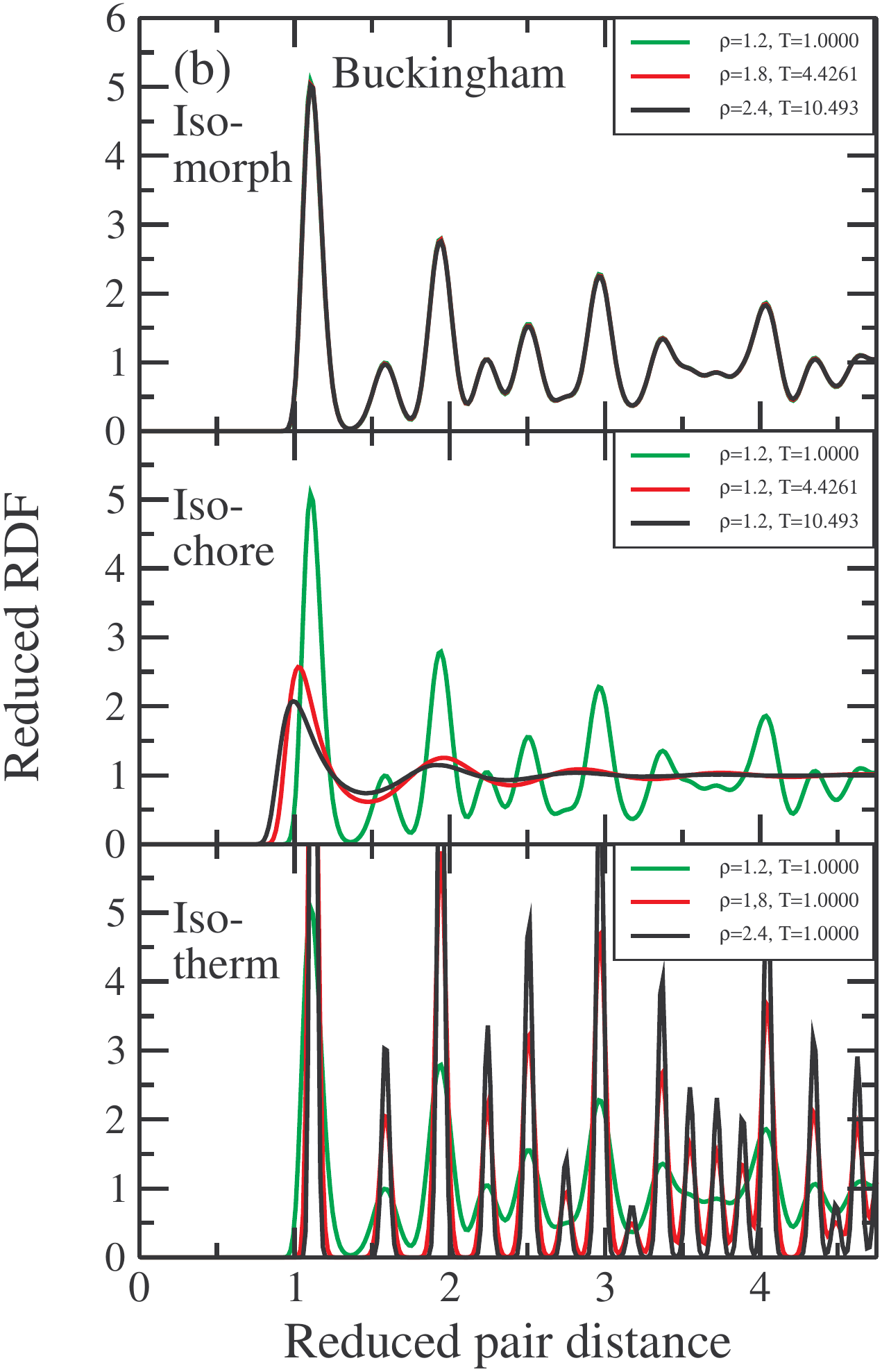}
\caption{(a) Direct isomorph check for the Buckingham potential FCC crystal. This system does not have a repulsive IPL term, but instead an exponentially repulsive term (Eq. (\ref{bpot})).
(b) RDFs along an isomorph and the corresponding isochore and isotherm, demonstrating isomorph invariance of the structure, much as for the LJ crystals.}
\label{Buck}
\end{figure}

Consider next whether crystalline isomorphs occur only for crystals of particles interacting via pair potentials consisting of IPL terms like the LJ potential. To investigate this we studied a FCC crystal composed of particles interacting via the Buckingham pair potential \cite{buc38}, 

\be\label{bpot}
v(r)=\varepsilon\left(\frac{6}{\alpha-6}\,e^{\alpha(1-r/\sigma)}-\frac{\alpha}{\alpha-6}\left(\frac{r}{\sigma}\right)^{-6}
\right)\,.
\ee
In contrast to the LJ case this potential's repulsive term is finite at $r=0$ (leading to a thermodynamic instability because the attractive term diverges at $r=0$, though for large values of $\alpha$ this is no problem in practice). The system simulated is that of $\alpha=14$ \cite{vel12}. In the unit system defined by $\sigma$ and $\varepsilon$, Fig. \ref{Buck}(a) shows the direct isomorph check for a doubling of the density starting from $\rho_1=1.2$ and $T_1=1$ from which we in the usual way calculated the temperature of the isomorphic state point at density $\rho_2=2.4$ ($T_2=10.493$). Figure \ref{Buck}(b) shows the RDFs for these two state points supplemented by a third isomorphic state point (upper subfigure). We conclude that for a crystal to have isomorphs it is not necessary that the pair potential is composed of IPL terms.

\begin{figure}
\centering
\includegraphics[width=.3\textwidth]{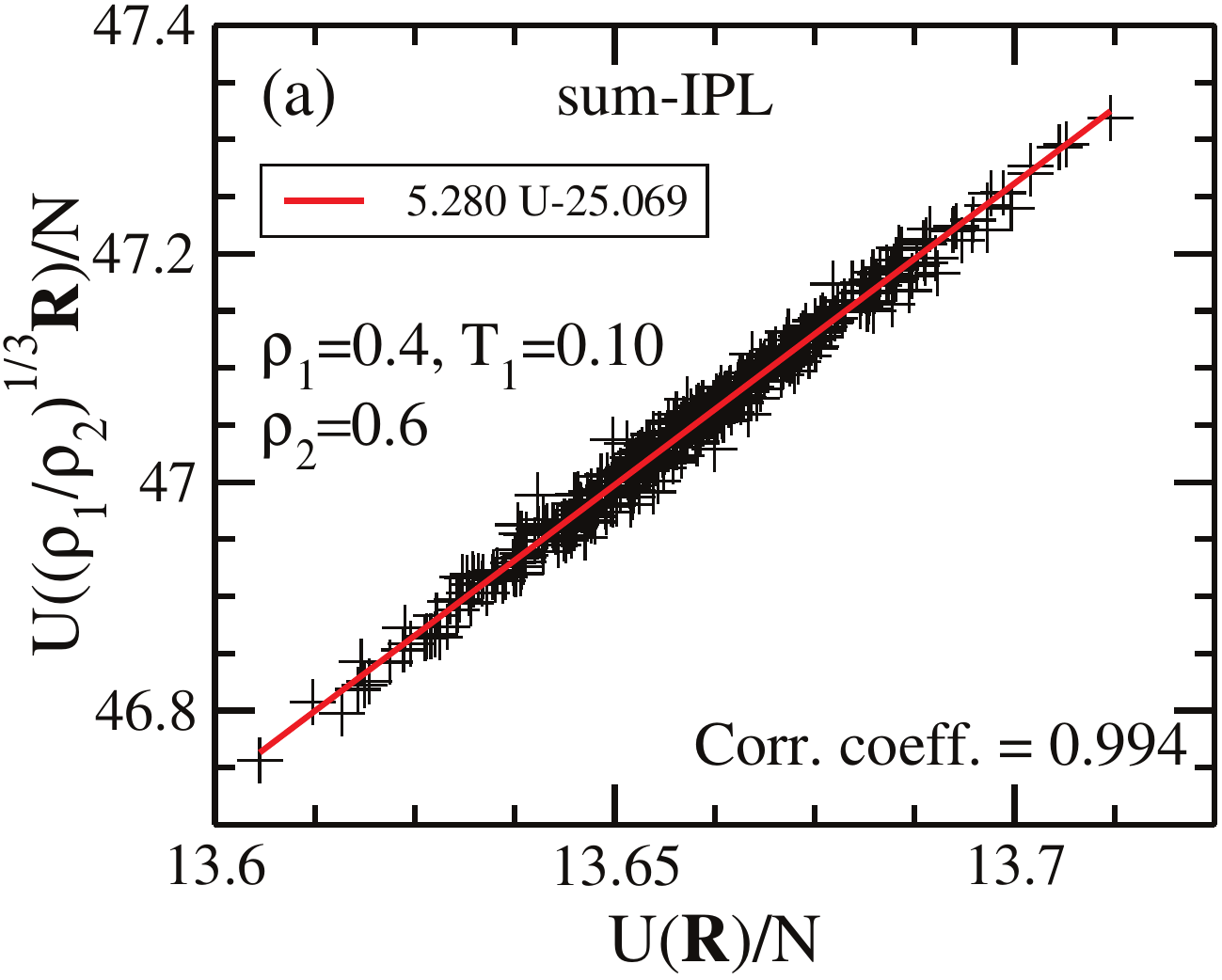}
\includegraphics[width=.3\textwidth]{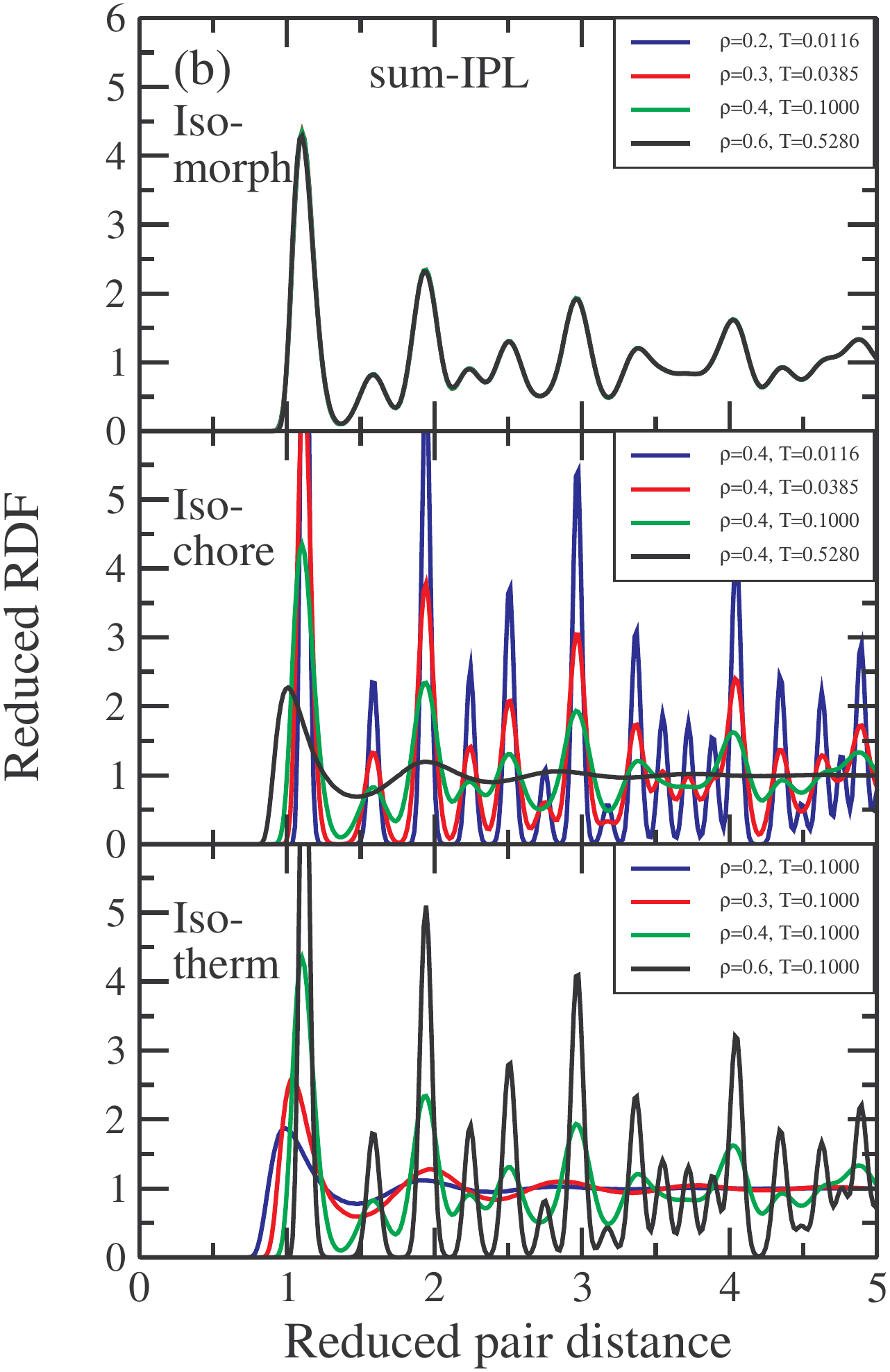}
\includegraphics[width=.3\textwidth]{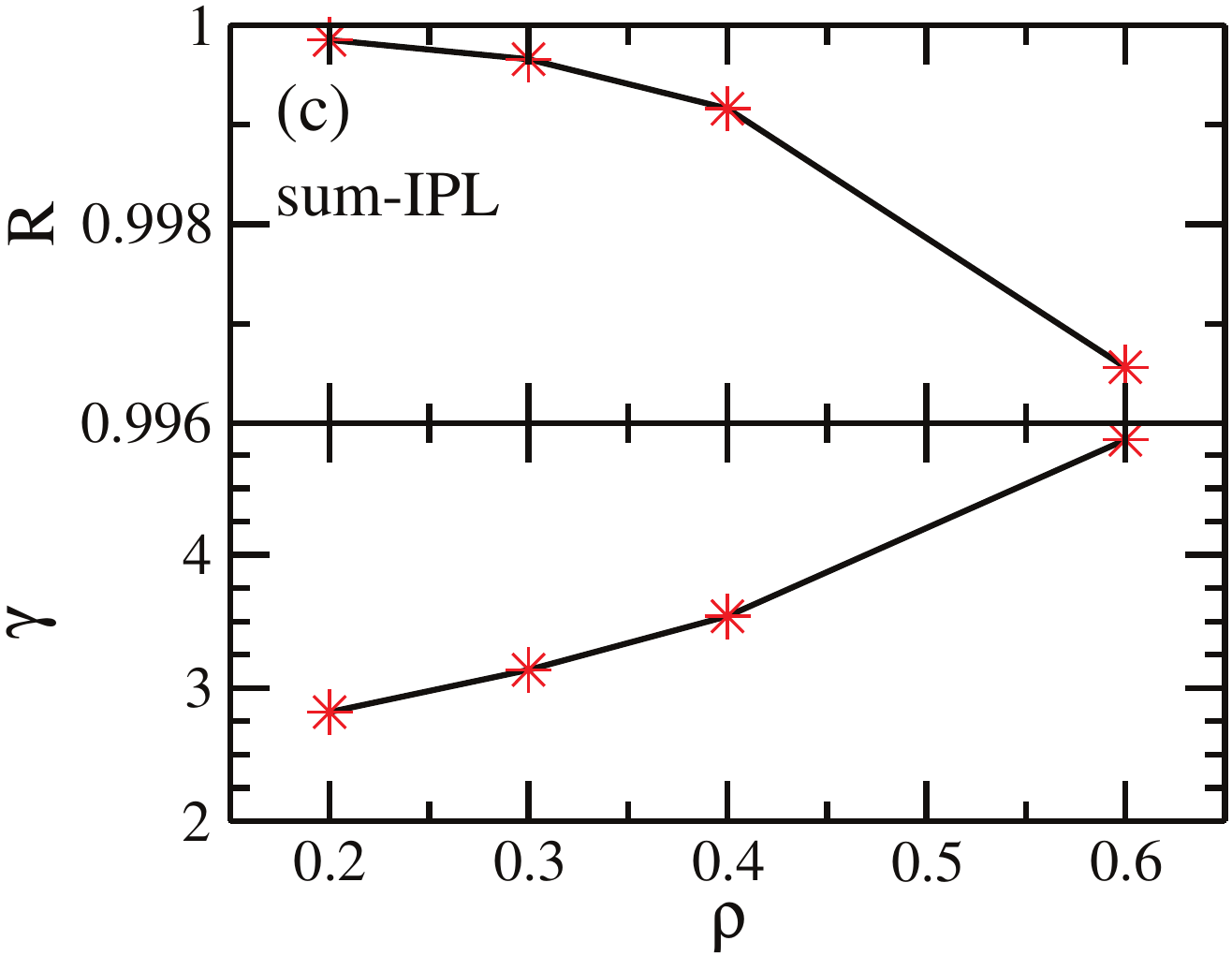}
\caption{(a) Direct isomorph check for an FCC crystal of particles interacting via the purely repulsive ``sum-IPL'' pair potential of Eq. (\ref{sumipleq}).
(b) RDFs along an isomorph, isochore, and isotherm, demonstrating isomorph invariance of the structure much as for the LJ systems.
(c) The viral potential-energy correlation coefficient (upper subfigure) and the so-called density-scaling exponent $\gamma=\langle\Delta W\Delta U\rangle/\langle(\Delta U)^2\rangle$ \cite{IV} (lower subfigure) as functions of density along the isomorph simulated in (b). The density-scaling exponent $\gamma$ is quite different from two, its value for an $r^{-6}$ IPL potential; this shows that the logarithmic term is important in the density range studied.}
 \label{sumipl}
\end{figure}

A pure IPL system has a homogeneous potential-energy function, 100\% virial potential-energy correlations, and perfect isomorphs \cite{IV}. Given that the LJ crystal has isomorphs and that the NaCl crystal does not, one may speculate that the latter system's ``problem'' is that it involves not just two IPL terms, but three. In order to test whether the number of IPL terms is crucial for how well isomorph invariance applies, we simulated the novel pair potential $v(r)=\varepsilon\int_{n_0}^\infty (r/\sigma)^{-n}dn$ that adds infinitely many IPL terms. Carrying out the integration for $n_0=6$, the parameter chosen for our simulations, leads to the following purely repulsive ``sum-IPL'' pair potential

\be\label{sumipleq}
v(r)\,=\,\varepsilon\,\frac{\,(r/\sigma)^{-6}}{\ln(r/\sigma)}\,\,\,\,(r>\sigma)\,.
\ee
This potential diverges at $r=\sigma$ and can only be studied at densities with average nearest-neighbor distance larger than $\sigma$. We simulated FCC crystals of densities ranging from $0.2$ to $0.6$ in the unit system defined by $\sigma$. The results are reported in  Fig. \ref{sumipl}, where (a) shows the direct isomorph check for a density change from $0.4$ to $0.6$. Figure \ref{sumipl}(b) shows the RDFs along the isomorph generated from this and two other direct isomorph checks also generated from $\rho=0.4$ simulations, covering a factor of three density variation (upper subfigure). The RDFs of the corresponding isochores and isotherms are shown in the lower subfigures. 

The sum-IPL system has excellent isomorphs. At low densities the logarithmic term is almost constant and the sum-IPL potential becomes dominated by the $r^{-6}$ term, i.e., approaches a single-IPL potential (that trivially has perfect isomorphs). To show that the $r^{-6}$ IPL term does not dominate at the densities studied, the lower subfigure of Fig. \ref{sumipl}(c) gives the density-scaling exponent $\gamma$ as a function of density for the four isomorphic state points of the upper subfigure of (b). For the $n=6$ IPL pair potential $\gamma=6/3=2$, so the logarithmic term is clearly important at the densities studied. Thus many power laws are indeed in play here, and we conclude that the presence of several IPL terms does not necessarily  imply poor isomorphs; this is not the NaCl system's ``problem''. -- The upper subfigure of Fig. \ref{sumipl}(c) shows the virial potential-energy correlation coefficient.

\begin{figure}
\centering
\includegraphics[width=.3\textwidth]{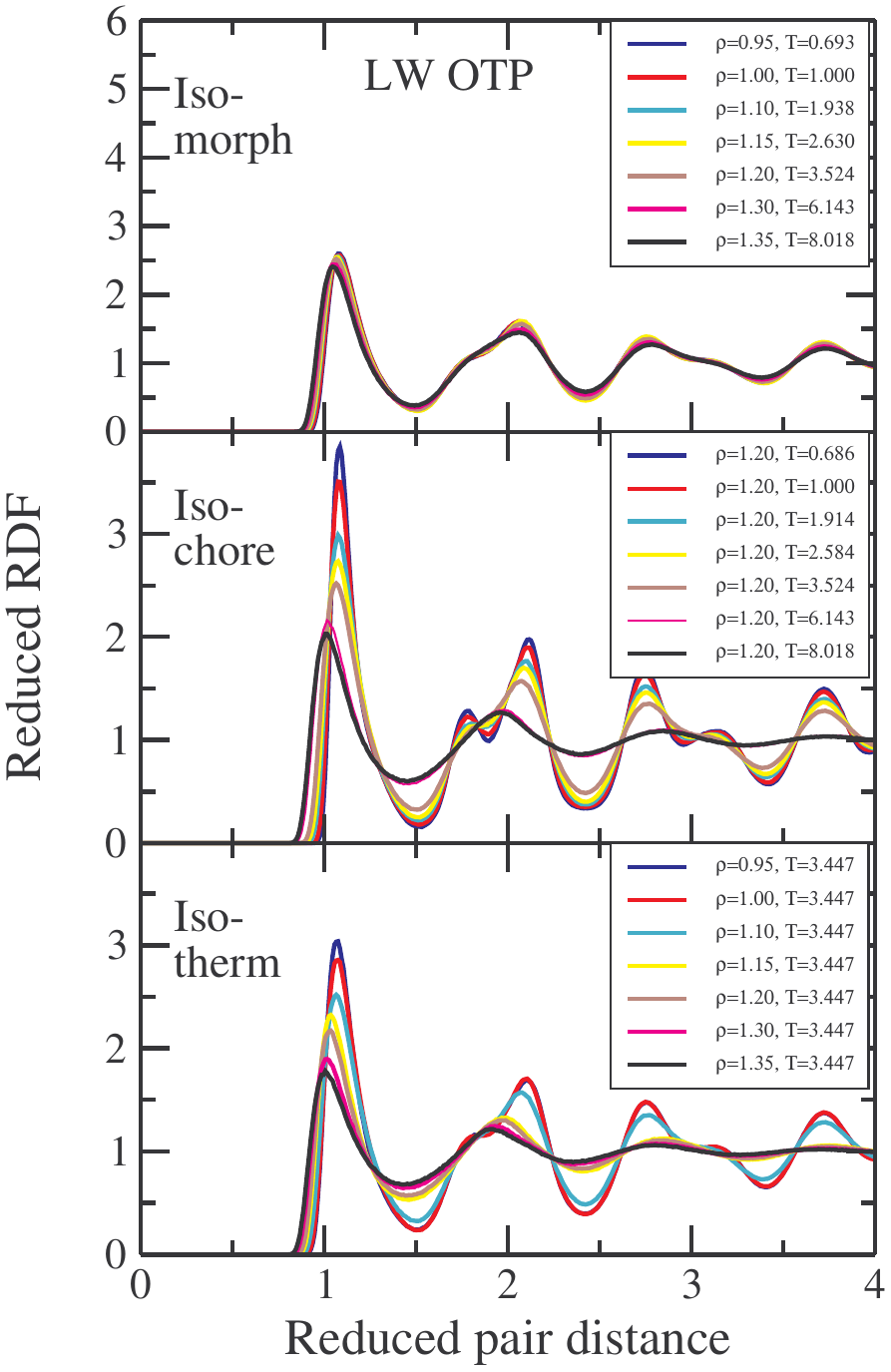}
\caption{RDFs along an isomorph and the corresponding isochore and isotherm of the Lewis-Wahnstr{\"o}m OTP crystal in which each molecule consists of three LJ spheres connected by rigid bonds with a $75^{\rm o}$  angle between the bonds (the RDFs refer to the LJ particles of different molecules). The isomorph was generated in the usual way from direct isomorph checks (not shown); the uniform scaling of the molecules for the direct isomorph checks generating the isomorphs keeps the bond lengths fixed. The figure demonstrates isomorph invariance of the structure, though not as accurate as for the LJ crystal.}
 \label{lwotp}
\end{figure}

The systems studied so far have been atomic crystals. The isomorph theory, however, is not limited to atomic systems. We end the paper by presenting results for two molecular crystals. First, we consider the Lewis-Wahnstr{\"o}m ortho-terphenyl (OTP) simple molecular model consisting of three LJ spheres connected by rigid bonds with angle $75^{\rm o}$ degrees \cite{otp2}. This model is difficult to crystallize and easily supercooled. In the crystal the three LJ spheres sit near the sites of a BCC lattice of a crystal with cubatic orientational disorder, i.e., with the molecules aligning randomly along the three Cartesian axis \cite{ped11a}. When a system like this in a simulation is scaled to a different density, the rigid bond lengths are kept constant and only the distances between the molecules' centers of masses are scaled uniformly to the new density. 

Figure \ref{lwotp} shows the simulation results for the intermolecular atom-atom RDF along an isomorph of the OTP crystal (upper subfigure) and along the corresponding isochore and isotherm. The isomorph was generated the usual way from direct isomorph checks (not shown). This time the density change is ``only'' 25\% and the collapse is not as good as for the LJ crystal (it is actually similar to that found for the LJ liquid). Nevertheless, this model still exhibits good data collapse along the isomorph. In this connection it should be kept in mind that typical high-pressure experiments often involve only a 5-10\% density change, so a change of 25\% is already quite large.

\begin{figure}
\centering
\includegraphics[width=.3\textwidth]{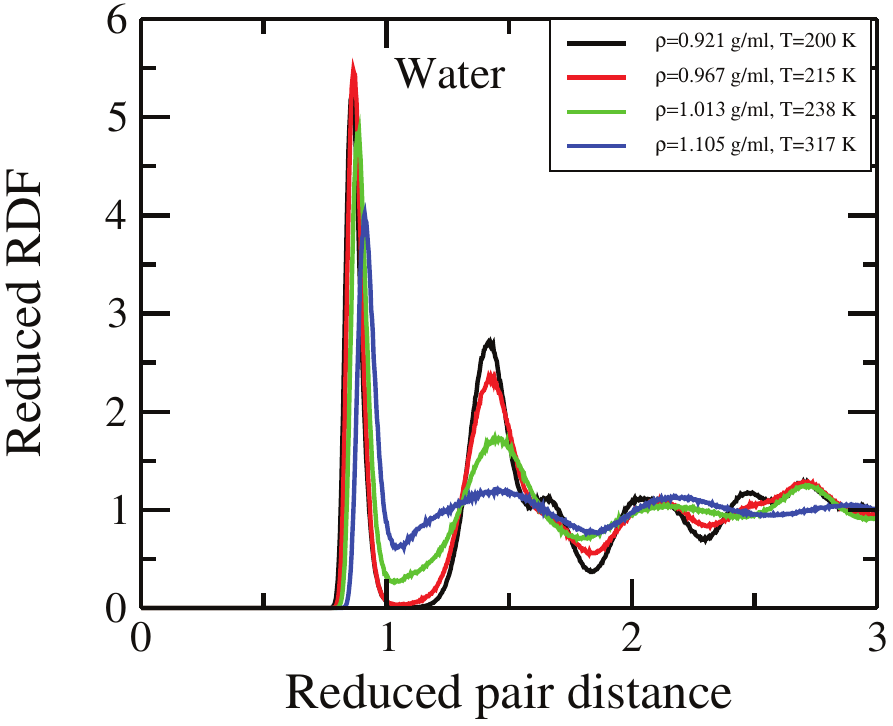}
\caption{Oxygen-oxygen RDF along a prospective isomorph generated from direct isomorph checks (not shown) for a 20\% density change of SPC/E hexagonal ice. The figure demonstrates that this crystal does not have isomorphs, a finding that is consistent with the fact that water is known to have almost zero virial potential-energy correlation coefficient \cite{I}.}
 \label{ice}
\end{figure}

In order to put the OTP results in perspective we finally show results for an ice model. Water has near zero virial potential-energy correlations at ambient conditions (a fact related to water's famous density maximum at $4^{\rm o}$ C \cite{I}), so ice is not expected to have isomorphs. We simulated the SPC/E water model \cite{spce} with the molecules arranged into a hexagonal crystal lattice. A  prospective isomorph was generated from a direct isomorph check plot (showing poor correlations). The result for the oxygen-atom RDF shown in Fig. \ref{ice} not surprisingly demonstrates that not all molecular crystals have isomorphs.

\section{Discussion}\label{disc}

We have studied the LJ crystal's structure and dynamics in some detail and demonstrated their reduced-unit invariance along isomorphs, as well as instantaneous equilibration following a jump between two isomorphic state points. The reduced-unit vibrational density of states is also an isomorph invariant, since it is determined by the velocity autocorrelation function \cite{dic69}. These findings validate hidden scale invariance for the crystalline LJ system as expressed in Eq. (\ref{prop1m}), a property that was previously studied only for liquids. The simulations show that the isomorph theory, in fact, works even better for the crystalline than for the liquid phase.
 
It is important to realize that the existence of isomorphs is not just a high-pressure phenomenon -- thus the lowest-density state point of the LJ crystal studied ($\rho=1.2$ and $T=1.0$) is close to the triple point ($\rho\cong 1.0$ and $T\cong 0.7$). The isomorph invariances actually continue to negative pressures (results not shown) as long as the system is metastable and does not phase separate into a crystal plus empty space, which happens around density 0.82 depending on the simulation length and system size.

Isomorph invariance does not apply for the NaCl crystal model. Thus the existence of isomorphs in crystals is not a trivial harmonic effect. Why does the NaCl crystal not have isomorphs? One difference between it and the LJ system is that NaCl is a two-component system.
Another difference is that the LJ system's pair potential involves only two IPL terms, whereas the NaCl system's involves three. Figures \ref{BiLJ} and \ref{sumipl} show, however, that neither fact explains the difference, which is probably due to the long-ranged nature of the NaCl system's strong Coulomb interactions, forces that have been shown to weaken the virial potential-energy correlations in the liquid phase \cite{I,sch09}.

For both the LJ and the NaCl systems the crystals and liquids behave in the same way as regards the existence of isomorphs. The same is the case for the five other models studied in Sec. \ref{other}. This is not trivial -- few non-universal properties survive the first-order transition separating the liquid and solid states of matter. Is it always the case that a liquid with isomorphs solidifies into a crystal that also has isomorphs? This is an open question, but our simulations so far indicate that the answer is probably yes because the crystal has always turned out to have stronger virial potential-energy correlations than the liquid. Of course, this still makes possible the interesting situation of a crystal with isomorphs that melts into a liquid without.

That the existence of isomorphs in crystals is not limited to the single-component LJ system's simple FCC crystal is illustrated by the other crystalline systems studied briefly in Sec. \ref{other}. In view of the simulation results of this paper and previous experiments and liquid-state simulations, in particular on supercooled liquids \cite{ing12}, there is good reason to believe that most or all metals and van der Waals bonded crystals have isomorphs and that this also applies for weakly ionic or dipolar crystals. On the other hand, covalently, hydrogen-bonded, and strongly ionic or dipolar crystals are not expected to have isomorphs. More work is needed to get a full overview of the situation, however. 

A recent reformulation of the isomorph theory \cite{sch14} throws new light on the condition for a crystal to have isomorphs. In the new formulation a Roskilde system is characterized by the property that the order of the potential energies of configurations at one density is maintained when these are scaled uniformly to a different density. If $\bRa$ and $\bRb$ are two physically relevant configurations at one density, this translates into the requirement $U(\bRa)<U(\bRb)\Rightarrow U(\lambda\bRa)<U(\lambda\bRb)$. As shown in Ref. \onlinecite{sch14} this condition implies all the fundamental characteristics of Roskilde systems \cite{IV,dyr14}, including the existence of isomorphs identical to the configurational adiabats, invariance of structure and dynamics along the isomorphs, and strong virial potential-energy correlations for constant-density fluctuations. To relate this new characterization of Roskilde systems to crystals, we assume that the crystal's potential energy is well described in the harmonic approximation. Adopting furthermore the Debye approximation, there are just two state-point dependent quantities characterizing the phonon spectrum, the transverse and longitudinal sound velocities. If the ratio between these is state-point independent, it is easy to see that the above inequality is obeyed because all phonon modes scale in the same way when density is changed. A more general formulation is the following: if the harmonic approximation applies and all phonon modes $\omega_{{\bf k},i}$ have the same Gr{\"u}neisen parameter $\partial\ln\omega_{{\bf k},i}/\partial\ln\rho$, the crystal is a Roskilde system.

It has been conjectured that at sufficiently high pressure all liquids have strong virial potential-energy correlations and thus isomorphs \cite{pap}. If this is confirmed, the same presumably applies for crystalline solids, implying that all crystals have isomorphs at sufficiently high pressure. This may explain the seemingly universal applicability of the high-pressure Gr{\"u}neisen equation of state (expressing a linear relation between pressure and energy with a constants that are a function only of the density) \cite{nag11}, which is equivalent to the thermodynamic separation identity characterizing a system with isomorphs, which in its configuration-space version has been shown to follow from Eq. (\ref{prop1m}) \cite{nag11,ing12a}.

Crystals with isomorphs have a thermodynamic phase diagram that is for many purposes effectively one- instead of two-dimensional. For a system with isomorphs the melting and freezing lines are themselves both isomorphs \cite{ped13,IV,V}. This is because the Boltzmann probabilities for crystalline configurations are isomorph invariant, implying that an isomorph cannot cross the melting or freezing lines. Note that this is a topological argument with no reference to free energies. As a consequence, crystals of Roskilde systems, i.e., systems with hidden scale invariance and obeying Eq. (\ref{prop1m}), are predicted to have constant excess entropy and invariant reduced-unit phonon spectra along the melting line. For such solids it is the excess entropy rather than the free energy that governs the reduced-unit structure and dynamics along the melting line. This fact resolves the apparent paradox mentioned in the Introduction of melting-line invariants referring only to a single phase (liquid or solid) \cite{nie11,gri12,lem13}. For instance, the Lindemann melting criterion -- according to which a crystal melts when the vibrational mean-square displacement reaches a certain fraction of the interatomic distance -- is predicted to be invariant along the melting line isomorph, i.e., melting takes place for the same reduced vibrational mean-square displacement. Such freezing/melting line invariances have been reported for LJ crystals \cite{lou05,cha07a} and in experiments \cite{ubb65}, though occasionally with some deviations at the lowest pressures. 

There are only few experimental studies of how the nucleation rate and mechanism and how the crystal growth rate and mechanism vary with pressure and temperature for a supercooled liquid  \cite{fre80,kel91,aue01}. The melting and freezing lines are isomorphs in parallel to a series of isomorphs in the coexistence as well as the supercooled liquid phases. Consequently, the isomorph theory predicts that nucleation and crystal growth properties depend only on the degree of supercooling as quantified by the excess entropy, not separately on pressure or temperature. This is predicted to apply for metallic, van der Waals bonded, and weakly ionic or dipolar supercooled liquids, but not for covalently bonded,  hydrogen-bonded, or strongly ionic or dipolar supercooled liquids. 

This paper has demonstrated that the existence of isomorphs is not limited to the standard, isotropic liquid state. One may ask whether isomorphs are present also in other anisotropic fluid states or, e.g., when a liquid is confined to certain geometries or subjected to external fields. Although more work is needed in this regard, there are strong indications that the answer is in the affirmative. For instance, simulations have shown that isomorphs exist for some liquids under nanoconfinement \cite{ing13a}, as well as  for liquids undergoing a linear or non-linear shear deformation described by the so-called SLLOD equations of motion \cite{sep13}. In reference to experiments, consider the transition between the nematic and isotropic phases of a liquid crystal. For a system with isomorphs, just as for melting, the nematic-isotropic transition line is an isomorph. Indeed, it has been shown for several liquid crystals that in the two-dimensional thermodynamic phase diagram the molecular ``flip-flop'' reorientational relaxation time is invariant along the nematic-isotropic phase transition line \cite{rol08a,urb11}. In other words, while the transition temperature varies with pressure along the transition line, the reorientational time does not. This is what one expects from isomorph invariance since the transition line is an isomorph along which the (reduced) relaxation time is consequently invariant.

At temperatures much below melting the structure, dynamics, and thermodynamics of a crystal become increasingly dominated by quantum effects. An important question for future work is whether there also in this region of the thermodynamic phase diagram are simplifying features for crystals of Roskilde systems, i.e, those that have classical-mechanical isomorphs at higher temperatures.

\begin{acknowledgments}
Ulf R. Pedersen gratefully acknowledges the support of the Villum Foundation's grant VKR-023455. The center for viscous liquid dynamics ``Glass and Time'' is sponsored by the Danish National Research Foundation via grant DNRF61.
\end{acknowledgments}

\end{document}